\documentclass[sigconf]{acmart}
\usepackage{balance}
\usepackage[utf8]{inputenc}%(only for the pdftex engine)
\def\BibTeX{{\rm B\kern-.05em{\sc i\kern-.025em b}\kern-.08emT\kern-.1667em\lower.7ex\hbox{E}\kern-.125emX}}

\setcopyright{none}

\copyrightyear{2021}
\acmYear{2021}
\acmConference[ACM IWSPA '21]{2021 ACM International Workshop on Security and Privacy Analytics}{April  28, 2021}{}
\acmBooktitle{2021 ACM International Workshop on Security and Privacy Analytics (IWSPA '19), Co-located with ACM CODASPY'21 April 26 --28}
\acmPrice{}
\acmDOI{}
\acmISBN{}
\usepackage{lineno}

\usepackage{longtable}
\usepackage{tabularx}
\usepackage{afterpage}
\usepackage{placeins}
\usepackage{rotating}
\usepackage{array}
\usepackage{todonotes}
\usepackage{xspace}
\usepackage[english]{babel}
\usepackage{soul}
\usepackage{flushend}
\usepackage{listings}
\usepackage{amsfonts}
\usepackage{amsmath}
\usepackage{graphicx}
\usepackage{subcaption}
\usepackage{multirow}

\usepackage{url}

\usepackage{breakurl}
\usepackage{cleveref}
\Crefformat{figure}{Fig.~#2#1#3}
\Crefmultiformat{figure}{Figs.~#2#1#3}{ and~#2#1#3}{, #2#1#3}{ and~#2#1#3}
\newcommand{\sysname}{\textsc{pricure}\xspace}

\makeatletter
\patchcmd{\@makecaption}
  {\scshape}
  {}
  {}
  {}

\newcommand*{\da@rightarrow}{\mathchar"0\hexnumber@\symAMSa 4B }
\newcommand*{\da@leftarrow}{\mathchar"0\hexnumber@\symAMSa 4C }
\newcommand*{\xdashrightarrow}[2][]{%
  \mathrel{%
    \mathpalette{\da@xarrow{#1}{#2}{}\da@rightarrow{\,}{}}{}%
  }%
}
\newcommand{\xdashleftarrow}[2][]{%
  \mathrel{%
    \mathpalette{\da@xarrow{#1}{#2}\da@leftarrow{}{}{\,}}{}%
  }%
}
\newcommand*{\da@xarrow}[7]{%
  \sbox0{$\ifx#7\scriptstyle\scriptscriptstyle\else\scriptstyle\fi#5#1#6\m@th$}%
  \sbox2{$\ifx#7\scriptstyle\scriptscriptstyle\else\scriptstyle\fi#5#2#6\m@th$}%
  \sbox4{$#7\dabar@\m@th$}%
  \dimen@=\wd0 %
  \ifdim\wd2 >\dimen@
    \dimen@=\wd2 %   
  \fi
  \count@=2 %
  \def\da@bars{\dabar@\dabar@}%
  \@whiledim\count@\wd4<\dimen@\do{%
    \advance\count@\@ne
    \expandafter\def\expandafter\da@bars\expandafter{%
      \da@bars
      \dabar@ 
    }%
  }%  
  \mathrel{#3}%
  \mathrel{%   
    \mathop{\da@bars}\limits
    \ifx\\#1\\%
    \else
      _{\copy0}%
    \fi
    \ifx\\#2\\%
    \else
      ^{\copy2}%
    \fi
  }%   
  \mathrel{#4}%
}
\makeatother

\newcolumntype{M}[1]{>{\centering\arraybackslash}p{#1}}

\usepackage{enumitem}
\setlist[itemize,1]{leftmargin=1.5\parindent, itemsep=0ex, topsep=0.5ex, % bottomsep=0.5ex
}
\setlist[enumerate,1]{leftmargin=2\parindent, itemsep=0ex, topsep=0.5ex, % bottomsep=0.5ex
}

\begin{document}

\fancyhead{}
  % do not delete this code.
 
  \title{\textbf{PRICURE}: Privacy-Preserving Collaborative Inference in a Multi-Party Setting}

\author{Ismat Jarin}
\affiliation{%
 \institution{University of Michigan, Dearborn}}
 \email{ijarin@umich.edu}

\author{Birhanu Eshete}
\affiliation{%
  \institution{University of Michigan, Dearborn}}
  \email{birhanu@umich.edu}

  \begin{abstract}
When multiple parties that deal with private data aim for a collaborative prediction task such as medical image classification, they are often constrained by data protection regulations and lack of trust among collaborating parties. If done in a privacy-preserving manner, predictive analytics can benefit from the collective prediction capability of multiple parties holding complementary datasets on the same machine learning task. This paper presents \sysname, a system that combines complementary strengths of secure multi-party computation (SMPC) and  differential privacy (DP) to enable privacy-preserving collaborative prediction among multiple model owners. SMPC enables secret-sharing of private models and client inputs with non-colluding secure servers to compute predictions without leaking model parameters and inputs. DP masks true prediction results via noisy aggregation so as to deter a semi-honest client who may mount membership inference attacks. We evaluate \sysname on neural networks across four datasets including benchmark medical image classification datasets. Our results suggest \sysname guarantees privacy for tens of model owners and clients with acceptable accuracy loss.  We also show that DP reduces membership inference attack exposure without hurting accuracy.

\end{abstract}
 
\settopmatter{printfolios=true}

\maketitle

\section{Introduction}\label{sec: intro}
Machine learning (ML) is being used in many application domains such as image classification~\cite{ImageNet}, voice recognition~\cite{DL-Speech2012}, medical diagnosis~\cite{DeepCC2019,MIMIC, IDC}, finance (e.g., credit risk assessment)~\cite{NN-Credit-Assess}, and autonomous driving \cite{DL-autnonmous17}. An emerging paradigm in ML is machine learning as a service (MLaaS) where clients submit inputs to obtain predictions via a cloud-based prediction API. When client inputs are privacy-sensitive (e.g., patient data), the MLaaS platform is expected to comply with privacy protection regulations (e.g., HIPAA in the United States) to preserve privacy of inputs (e.g., patient diagnosis details). Moreover, clients may not be encouraged to submit inputs that, if revealed to others may jeopardize their competitive advantage (e.g., on inputs about intellectual property).

In a collaborative setting where multiple MLaaS providers own private models trained on their respective private data, and clients own private input samples, clients can benefit from the collective inference capability of multiple model owners for tasks such as medical image classification. As a result, a practical setting for multiple MLaaS providers to collaborate on a common ML task is to keep the secrecy of their respective models trained on private data and participate in a privacy-preserving common predictive task such that: after a single iteration of collaborative inference, $(a)$ the client learns nothing about the models of MLaaS providers; $(b)$ the MLaaS providers learn nothing about client’s data, and $(c)$ the inference capability of a semi-honest client is limited.

To build privacy-preserving mechanisms in to the ML pipeline, previous work proposed private training methods based on objective perturbation \cite{DP-Empirical-RM11}, gradient perturbation \cite{DP-SGD16,DistribLearning18}, and output perturbation \cite{DP-SGD16,DistribLearning18}. In the collaborative setting, prior work has proposed cryptography-based transformation of ML building blocks (e.g., activation functions, pooling operations) at training time (e.g., CryptoNets~\cite{CryptoNets16}, SecureML~\cite{SecureML17}, \cite{DistribLearning18}), oblivious transformation of neural networks (e.g., MiniONN~\cite{MiniONN17}), and secret sharing (e.g., Chameleon~\cite{Chameleon}, SecureNN~\cite{SecureNN}) to enable 2- and 3-party secure computation for collaborative training and inference. However, no prior work explores the collaborative inference setting with tens of model owners. Moreover, when a client is semi-honest, it might take the oracle-style interaction with the MLaaS provider to mount membership inference style attacks.

In this paper, we present \sysname\footnote{PRICURE stands for \underline{PRI}vacy-preserving \underline{C}ollaborative inference in a m\underline{U}lti-pa\underline{R}ty s\underline{E}tting.}, a system that combines complementary privacy protection notions of {\em secure multi-party computation (SMPC)} and {\em differential privacy (DP)} to enable privacy-preserving collaborative inference among multiple model owners holding private models and clients holding private input samples. The key insight behind combining SMPC and DP is the orthogonal protections they provide. Intuitively, given an input $x$ and a function $f(x)$, SMPC is aimed at avoiding information about $x$ that is leaked in the course of computing $f(x)$. DP, on the other hand, aims to randomize $f$ such that an adversary has very limited leverage to infer about $x$ by observing $f(x)$. In the collaborative inference we consider in \sysname, SMPC addresses { \em pre-inference disclosure protection of private data} while DP addresses {\em post-inference protection of inference results} to limit attacks such as membership inference. This orthogonal nature of SMPC and DP makes their combination appealing for our setting. 

The privacy guarantee in \sysname stems from the notion of {\em additive secret sharing} \cite{SecureNN, SPDZ_mult}, where model owners train their private models and secret-share model parameters with non-colluding secure servers (we call them {\em workers}) that compute intermediate results on secret-shared input samples private to clients and similarly secret-shared by a client with the workers. Using intermediate inference results from the workers, a {\em trusted aggregator} reconstructs the final inference results and performs noisy aggregation to deter a semi-honest client who may mount membership inference attacks. 
While we borrow the additive secret sharing notion from prior work (SecureNN~\cite{SecureNN}, SPDZ~\cite{SPDZ_mult}) which focus on privacy-preserving training and inference in a 3-party setting, in \sysname we rather focus on secret-sharing-based privacy-preserving collaborative inference to handle tens of model owners with acceptable accuracy/privacy trade-off, while also  providing a differential privacy-based layer of defense against membership inference attacks that exploit the oracle access to the MLaaS provider. 

We evaluate \sysname on neural networks across four datasets that span handwritten digit recognition (MNIST \cite{MNIST}), clothing image classification (Fashion-MNNIST \cite{Fashion-MNIST}), breast cancer classification (IDC \cite{IDC}), and in-ICU patient length-of-stay prediction (MIMIC \cite{MIMIC}). For instance, on the MNIST dataset, using a commodity hardware setup, our results guarantee privacy for up to 50 model owners with nearly no accuracy loss, with a per-model average overhead of 48ms for secret-sharing model parameters, and a per-model response delay of 1.5s. On the MIMIC dataset, we show that DP reduces the  accuracy of membership inference attack by up to 9.02\%, which demonstrates the utility of differently privacy as a second layer of protection against inference on top of the SMPC-based disclosure protection for inputs and model parameters. In summary, we make the following contributions:

% \begin{itemize}
$\bullet$ \textbf{New framework:} Through novel combination of orthogonal privacy guarantees of secure multi-party computation and differential privacy, we use additive secret sharing to preserve privacy of model parameters for model owners and input samples for clients; and we provide a differential privacy-based layer of defense against membership inference attack so as to limit adversarial advances of a semi-honest client.

$\bullet$ \textbf{Scalable approach:} We build on prior work \cite{SecureNN,SPDZ_mult} and demonstrate a scalable collaborative inference with tens of model owners with acceptable accuracy/privacy trade-off.

$\bullet$ \textbf{Comprehensive evaluations:} We conduct extensive experiments with four datasets (MNIST \cite{MNIST}, Fashion-MNIST \cite{Fashion-MNIST}, IDC \cite{IDC}, and MIMIC \cite{MIMIC}) that include medical datasets, with varying number of model owners, and across a range of privacy budget values to evaluate: accuracy/privacy trade-off, impact of number of model owners on accuracy/privacy, resilience against membership inference attack, and performance overhead of \sysname overall, per-sample, and per-model-owner.

$\bullet$ \textbf{Our code is available at}:  \texttt{ \url{https://github.com/um-dsp/PRICURE}}.
\section{Background and Preliminaries}\label{sec: bground}
In this section, we cover preliminaries focusing on neural networks, secure multi-party computation, and differential privacy. 

\subsection{Feed-Forward Neural Networks}
A feed-forward neural network (FFNN) is a network of information processing units called neurons organized into layers. In a FFNN, information travels only forward without looping, starting from input nodes, then through hidden nodes, and finally to output nodes. The goal of the FFNN is to approximate some function $F_{\theta}$, and then map an input $x$ to a label $y$ as $y=F_{\theta}(x)$, where $\theta$ is the learned parameter vector. 

The network architecture consists of the following: an input layer that captures raw inputs, hidden layers that apply a series of transformations to the input,  and the output layer that produces the mapping of $x$ to $y$. For a multi-class classifier, the output layer has as many neurons as the class labels. The coefficients of connections between two neurons are referred to as {\em weight}, and are typically initialized as small random values to bootstrap the training process. 

Suppose that the FFNN has $l$ number of hidden layers, and we denote the number of neurons at layer $l$ as $k_l$. The output layer has $o$ number of neurons. The input vector ($x_1, x_2, ..., x_d$) is the $d$-dimensional input to the network. Without loss of generality, the label $y$ is computed as:  
%\begin{align}\label{eq:basicNN}
$y = F(x,\theta)=\sum_{j=0}^{k} \xi_j\theta_j$,
%\end{align}
 where $\theta$ are model parameters, $\xi$ is a vector of non-linear parametric basis functions where $\xi_0(x)=1$. %The $j^{th}$ basis function can be represented as:
%\begin{align} \label{eq:background_basis}
%\xi_j(x)&=h(\sum_{i=1}^{d} W_{ij} x_i +b_j)
%\end{align}
Now, given $k_1$ number of neurons in the $1^{st}$  hidden layer, the output of $j^{th}$ neuron of the $1^{st}$ hidden layer is computed as: 
%\begin{align} \label{eq:1st_hidden_layer}
$r^1_j(x,W)=h(\sum_{i=1}^{d} W_{ij}(1) x_i +b_j)$,
%\end{align}
 where $W_{ij}(1)$ is the weight from connection between the input to connection of the $1^{st}$ hidden layer of $j^{th}$ neuron, $b_j$ is the bias and $h$ is the non-linear differentiable activation function. As a result, for the $1^{st}$  hidden layer, we have the output neurons $[r^1_1,r^1_2,....,r^1_{k_1}]$. These output neurons will then be the input vectors to the $2^{nd}$  hidden layer. In the same vein, the output of the $j^{th}$ neuron from the $l^{th}$ hidden layer is computed as:
%\begin{align*} %\label{eq:hidden_layer}
%r^2_j(x,W)&=h(\sum_{i=1}^{k_1} W_{ij}(2) r^1_i )\\
$r^l_j(x,W)=h(\sum_{i=1}^{k_{l-1}} W_{ij}(l) r^{l-1}_i )$.
%\end{align*}
Finally, the result of the $j^{th}$ neuron in the output layer is computed as: 
%\begin{align}\label{eq:outNN}
$y_j(x,W)=h(\sum_{i=1}^{k_{l}} W_{ij}(l+1) r^l_i )$,
%\end{align}
 where the coefficient $W_{ij}(l+1)$ is the weight vector from the final hidden layer $l$ to the $j^{th}$ node of the output layer. The resultant vector for all neurons of the output layer will be $[y_1, y_2,..., y_o]$. Note that, the weight vector $W(1) \in R^{(d \dots k_1)}$, $W(2) \in R^{(K_1 \dots k_2)}$, ...,  $W(l+1) \in R^{(k_l \dots o)}$, are concatenated into the parameter vector $\theta =[W(1), ..., W(l+1)]$.

\subsection{Secure Multi-Party Computation}
Secure multi-party computation (SMPC)~\cite{SMPC} enables a set of parties $P = \{P_1,...,P_m\}$, where party $P_i$ holds sensitive input $x_i$, to jointly compute a function $y = f(x_1,...,x_m)$ while protecting each $x_i$. The computation needs to result in the correct value of $y$ (called the {\em correctness property}) and at the end of the computation each $P_i$ learns nothing beyond $y$ (called the {\em privacy property}). SMPC has several applications, for example in:  privacy-preserving decision making on distributed data~\cite{Decision_Making}, privacy-preserving machine learning~\cite{CryptoNets16}, secure auctions~\cite{Secure_Auction}, and secure voting~\cite{Secure_Voting}.

The two popular implementations of SMPC are {\em garbled circuit}~\cite{GC} and {\em secret sharing}~\cite{Secret_share1}. In garbled circuit, the $P_i$'s construct a (large, encrypted) circuit and evaluate it at once, while in secret sharing they need to interact for each circuit gate. Garbled circuit allows for constant number of rounds but requires larger bandwidth (i.e., fewer messages, but bigger messages are sent). Secret sharing has typically low bandwidth (i.e., small messages per gate) and high throughput, where the number of rounds is determined by the depth of the circuit. In this work, we use secret-sharing-based MPC. Intuitively, a $(t,n)$-secret sharing scheme splits a secret $s$ into $n$ shares $s_i$ and at least $t$ shares are required to reconstruct the secret. We use $\langle s \rangle = (s_1,...,s_n)$ to indicate the sharing of $s$ among $n$ parties.

{\em Additive secret sharing}~\cite{Secret_share2} allows a secret $s$ to be split into random parts and shares them with {\em secure workers} (real/virtual instances that perform computations such as addition and multiplication securely). For example, suppose there are two workers $A$ and $B$ and a secret $s$. The above notation then becomes $\langle s \rangle = (s_1,s_2)$ because $n=2$. Accordingly,  $A$ and $B$ receive share values $s_1^a$ and $s_2^b$, respectively. The workers $A$ and $B$ perform computation (e.g., compute function $F$) directly on the share values. After finishing the computation, the workers produce intermediate results as follows: 
\vspace{-.085in}
\begin{align}
A: y_a & =F(s_1^a)\\
B: y_b & =F(s_2^b)
\end{align}
These intermediate results $y_a$ and $y_b$ are then combined using {\em private additive scheme} and then the true output result is revealed. This secret sharing operation does not use floating point numbers, rather performed in a mathematical space called {\em Integer Quotient Ring}~\cite{Q_ring}, which contains the integer between 0 to Q-1. Here Q is a prime number that has to be big enough so that the space is able to contain all the numbers that would be used in our experiments. A conceptual proof that illustrates additive secret sharing is described in the Appendix (Section \ref{subsec: secret-sharing-proof}). 

Our implementation of \sysname builds on private additive sharing introduced in SecureNN~\cite{SecureNN_1} and incorporated in  PySyft \cite{PySyft}. SecureNN builds on MPC to implement exact non-linear functions while avoiding the use of inter-conversion protocols as well as general-purpose number-theoretic libraries. The mathematical operations like matrix multiplication, summation, private comparison, division, and max-pooling operations are built based the MPC property. PySyft implements SPDZ \cite{SPDZ_protocol}, a secret sharing scheme that enables more complex operations. In the Appendix (Section \ref{subsec: spdz-proof}), we provide an illustrative proof on how SPDZ functions.

\subsection{Differential Privacy}
For two neighboring datasets $D_i$ and $D_j$, which differ in just one data point $x$, suppose $F_i$  is trained on $D_i$ and $F_j$  is trained on $D_j$. Let the output space of $F_i$ and $F_j$ be $S$ such that for an input $x$, $F_i(x) \in S$ and $F_j(x) \in S$. Differential privacy (DP)~\cite{Differential_Privacy} guarantees that a randomized mechanism $M(F_i(x))$ and $M(F_j(x))$ does not enable an observer (adversary) to distinguish whether $M$'s output was based on $D_i$ or $D_j$, i.e., whether or not $x$ is used as a training example for $F_i(x)$ or $F_j(x)$. The indistinguishability of $x$'s membership in $D_i$ or $D_j$ protects the identifiability of $x$ (e.g., a person's medical record). In $\epsilon$-DP, the indistinguishability of the outputs of $F_i(x)$ and $F_j(x)$ is parametrized by $\epsilon$ (also called the privacy budget). Equation ~\ref{eq:dp} formalizes the notion of ($\epsilon$)-DP as follows:
 \begin{equation} \label{eq:dp}
P[M(F_i(x)) \in S] \le e^\epsilon \times P[M(F_j(x)) \in S]
\end{equation}

Intuitively, lower $\epsilon$ values indicate stronger privacy protection.  Equation \ref{eq:dp} has the $(\epsilon, \delta)$ variant: $P[M(F_i(x)) \in S] \le e^\epsilon \times P[M(F_j(x)) \in S]+ \delta$, where $\delta$ refers to the failure probability of the mechanism M, and when $\delta =0$, we say $M$ is $\epsilon$-DP. The $\epsilon$-DP formalism in Equation~\ref{eq:dp} guarantees individual data item privacy in the most extreme case where  $D_i$ and $D_j$ are so similar that only one data point sets them apart. In \sysname, each $D_i$ is unique since the datasets come from independent data owners. Hence, using the DP notion, \sysname guarantees that individuals who contribute privacy-sensitive data have bounded privacy guarantee.

In DP, randomization is at the core of ensuring the indistinguishability of an individual’s record in a dataset. A typical way to achieve output randomization is to add {\em noise} to an output (e.g., $F_i(x)$) using, for instance, the {\em Laplace mechanism}. For a privacy budget $\epsilon$ and $F_i$’s sensitivity value of $s$, using the Laplace mechanism $Lap(b)$ centered at $0$ and scale $b$, $noise$ is computed as $Lap(\frac{s}{\epsilon})$. 

\section{Problem Statement and Threat Model}\label{sec: prob-stmt}
In this section, we discuss our problem statement and threat model with respect to the parties that take part in \sysname.
\subsection{Problem Statement}
\textbf{Problem:} We consider a setting where multiple model owners $P= \{P_1, …, P_m\}$ hold private models $F_1, …, F_m$ such that $F_i$ is trained on $P_i$’s private data, and client $C$ submits their private input $x = [x_1, …, x_d]$ to obtain an inference result:
% \begin{equation}
$y = \mathbf{\Phi}_{i: 1...m}(F_i(x))$,
% \end{equation}
where $\mathbf{\Phi}$ is an aggregation function that leverages the collective inference capability of $F_i$'s. Taking a feed-forward neural network $F_i$ with number of layers $l$, the collaborative inference result can be expanded in terms of weights and biases as:\\
% \begin{equation}
    $y =  \mathbf{\Phi}_{i: 1...m}(W_l \cdot F^i_{l-1}(...F^i_1 (W_1 \cdot x + b_1)...) + b_l)$.
% \end{equation}

\textbf{Goals:} Our first goal is to obtain $y$ while keeping $F_i$'s and $x$ private. In particular, after each prediction, $C$ learns nothing about $F_i$’s (i.e., $W_1, W_2, ..., W_l$ and $b_1, b_2, ..., b_l$) and $P_i$’s learn nothing about $x$ and other model owners. Our second goal is to deter a semi-honest client $C$ from mounting membership inference attacks.

\textbf{Challenges:} To achieve the aforementioned goals, one needs to address specific research questions with regards to a) accuracy/privacy trade-off, b) scalability with growing number of model owners, c) the prospect of attacks such as membership inference even in the presence of privacy-preserving schemes, and d) performance implications of privacy-enhancing schemes in a multi-party setting. We address these questions in Sections \ref{subsec: acc-priv-trade-off} - \ref{subsec: preformance-overhead}.

\subsection{Threat Model}
% What are the goals of the adversary?
% What is the knowledge of the adversary?
% What are the capabilities of the adversary?
Here, we describe our threat model with respect to model owners, workers, aggregator, and client.

\textbf{Model Owners:} We consider the {\em semi-honest} setting for model owners, whereby they do not trust each other to share their training data and/or models, because of data protection and privacy regulations (e.g., HIPAA) or competitive advantage (e.g., they compete in the same business).

\textbf{Client:} The client is assumed to be {\em semi-honest} for it may use \sysname as an oracle to initiate membership inference~\cite{MIA}, model extraction~\cite{model-stealing16}, or model inversion attacks~\cite{model-inversion}. Moreover, it does not have access to model parameters of any of the model owners.

\textbf{Workers:} In the {\em honest-but-curious} sense, they may analyze the secret-shares they receive from model owners or intermediate computation results, but are trusted enough not to collude with each other to exchange their respective secret-shares or intermediate computation results. Given the partial result they compute, workers are assumed to have no access to the final inference result.

\textbf{Aggregator:} We assume the aggregator is {\em trusted by model owners and clients} not to reveal true inference results to any third party, and will not mount model extraction/membership inference style attacks. Moreover, it doesn't have access to model parameters of the model owners and the data of the client. When sending the prediction results to the client, the aggregator encrypts the result with the client's public key (shared with the aggregator a priori).

\section{Approach}\label{sec: approach}
In this section, we first give a high-level overview of \sysname and discuss details in Sections \ref{subsec: model-owner} -- \ref{subsec: aggragator}.

\subsection{Overview}\label{subsec:approach-overview}

\begin{figure*}[t!]
\begin{centering}
    \includegraphics[width =.99\textwidth]{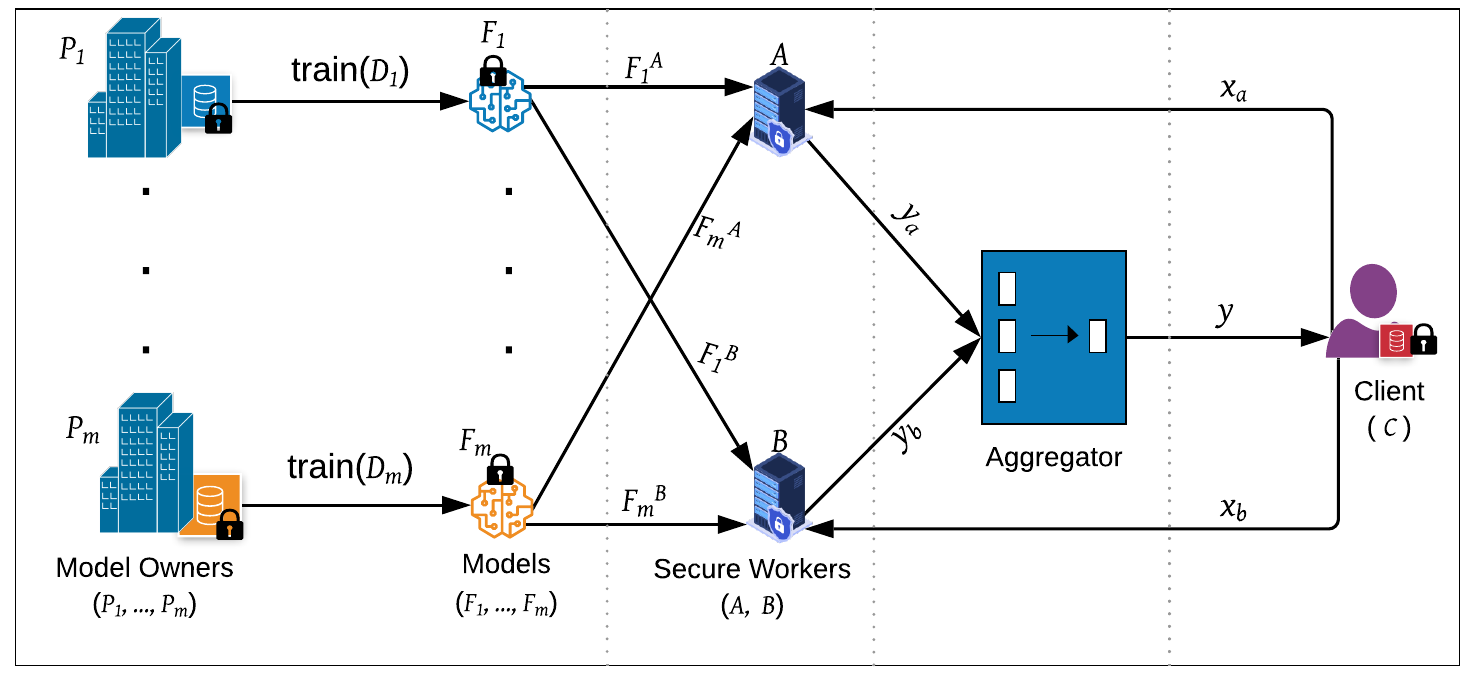}
\end{centering}
\caption{\sysname Overview. {\small Model owners $P_1, ..., P_m$ train private models $F_1, ..., F_m$. The $P_i$'s then secret-share $F_i$'s with workers $A$ and $B$ as $<F_{i}> = (F_i^{A}, F_i^{B})$. A client $C$ secret-shares private sample $x$ with $A$ and $B$ as $<x> = (x_a,  x_b)$. Using secret-shares of models $F_1$ to $F_m$, $A$ and $B$ each compute intermediate results as $y_i^{a}= F_i^{A}(x_a)$ and $y_i^{b} = F_i^{B}(x_b)$, respectively, and send them to the aggregator. The aggregator reconstructs inference result $y_i$ from $y_i^{a}$ and $y_i^{b}$. Finally, the aggregator releases to the client a differentially private aggregation of the $y_i$'s as $y$.}}
\label{fig:framework}
\end{figure*}

Figure \ref{fig:framework} highlights our proposed system, \sysname. We consider a setting of $m$ model owners (e.g., hospitals) who may not share training data and models due to regulatory or trust reasons, yet they want to participate in a collaborative inference task $T$ (e.g., medical image classification). As a result, model owners train $m$ private models $F_1$, ..., $F_m$. \sysname combines {\em secure multi-party computation (SMPC)} and {\em differential privacy (DP)} to enable collaborative inference among the $m$ model owners on task $T$, while $(a)$ ensuring secrecy of each $F_i$ and samples submitted for inference from client $C$ and $(b)$ limiting adversarial advances of a semi-honest client to mount membership inference attack.

\textbf{Why Combine SMPC and DP?} It is because these two schemes are complementary to each other, and they provide orthogonal protections to address $(a)$ and $(b)$. Given an input $x$ and a function $f(x)$, SMPC is aimed at answering ``what information about $x$ is leaked in the course of computing $f(x)$?’', which addresses $(a)$. DP, on the other hand, is concerned with ``what can be inferred by analyzing $f(x)$?’’, which addresses $(b)$. In the sense of the collaborative inference we consider in \sysname, SMPC is relevant for {\em pre-inference protection} of data while DP is aimed at {\em post-inference protection} to avoid attacks such as membership inference.  

\vspace{-.2em}

After training a model on a private dataset, each model owner secret-shares its model parameters with non-colluding secure servers, which we call {\em workers}. A client $C$ (e.g., another hospital, a medical researcher) who wants to obtain an inference result (e.g., classification) on sample $x$ also secret-shares the sample with the workers. A single iteration of collaborative inference proceeds as follows:  when the workers receive a secret-share of an input $x$, they use each model’s partial inference function to compute and return a partial result to a {\em trusted aggregator}, which reconstructs each model’s partial inference results into full inference output for sample $x$. The aggregator then performs confidence-weighted aggregation of the inference results, and finally leverages differential privacy to add random noise to the final inference result and sends it to the client.

The secrecy of client's inputs and model parameters of each owner is protected using {\em additive secret sharing}. Noisy inference via $\epsilon$-DP is used to deter a semi-honest client who might attempt to mount membership inference attacks. The communication link between the client and the aggregator is a secure channel. The client shares its public key with the aggregator, that it uses to encrypt an inference result so the client decrypts it with its private key to obtain the final output of the collaborative inference for sample $x$.

\subsection{Model Owners}\label{subsec: model-owner}
 Each model owner maintains its private training data from which it trains a private model, and does not share both training data and model details with other model owners, the aggregator, or a third party. To enable collaborative inference that benefits from the collective intelligence of what is learned by each $F_i$, model owners are, in principle, free to use any model architecture, but need to agree a priori on a common feature representation (e.g., pixel intensities for images) and the inference output format (e.g., label only, probability score). A model owner $P_i$ may train its respective model $F_i$, with or without privacy. We note that when a model is trained in a privacy-preserving manner (e.g., via gradient perturbation as in differentially private stochastic gradient descent (DP-SGD)~\cite{DP-SGD16} or using output perturbation \cite{DistribLearning18}), the privacy budget will add up with the inference output perturbation \sysname introduces at the aggregator, which may incur more penalty on accuracy.

\textbf{Number of Model Owners:} The minimum value for $m$ is 2 while there is no limit on the maximum. Intuitively, the higher the value of $m$, the better the collective inference accuracy of the participants. This, however, depends on the accuracy of each model and the privacy budget (amount of noise added on the final inference output). Our model owner setup shares some similarity with  PATE~\cite{PATE17}. However, our setting differs both in the goals and the details. For instance, in PATE, the larger $m$ implies potentially better privacy guarantee. In our case, the interpretation of how large $m$ may not inherently entail better privacy guarantee, since each model owner is acting on its own training data (unlike PATE, where training data is partitioned into $m$ disjoint sets). The disjoint $m$ partitions in PATE is equivalent to the typically disjoint $m$ model owners in \sysname. Model owners in \sysname may use varying size of training data and the choice of $m$ may have implications on the accuracy/privacy trade-off (in PATE, depending on the dataset size, the choice of $m$ may affect accuracy/privacy trade-off). In Section \ref{subsec: model-ownwers-vs-acc-priv}, we will analyze the relationship between number of model owners, inference accuracy, and privacy guarantee. Next, we describe how model owners secret-share model parameters with workers.

\textbf{Secret-Sharing of Model Parameters:} Model owners secret-share their model parameters with the virtual workers via a secure channel. Once they secret-share their model parameters, model owners are not required to stay connected with the workers. Instead, they can terminate the secure connection and initiate it later as needed (for instance when a model is retrained and updating the model parameters is deemed necessary).

Once a model owner trains a model, the learned model parameters are captured via weights ($W$) and bias ($b$). Suppose for model owner $P_i$  ($P\in [1,m]$), the weight vectors are $\theta=[W(1),....,W(l+1)]$, where, the weight vector $W(1)$ represents the weight parameters from input to the $1^{st}$ hidden layer, similarly $W(2)$ represents the weight parameters from $1^{st}$ hidden layer to the $2^{nd}$ hidden layer and $W(l+1)$ represents the weight parameters from $l^{th}$ hidden layer to the output layer. Using Equation \ref{eq:share-a} and \ref{eq:share-b}, model owner $P_i$ secret-shares its $j^{th}$ model parameters with worker $A$ and $B$, respectively as follows:
\begin{align*} %\label{eq:o1}
   Worker A:W^a(j)&=Random(-Q,Q) \\
   Worker B: W^b(j)&= W(j)-W^a(j)
\end{align*}
Finally, all secret-shared model parameters are computed as follows:
\begin{align*} %\label{eq:ow1}
   Worker A: \theta^a=[W^a(1),.....,W^a(l+1)],b_i^a \\
   Worker B: \theta^b=[W^b(1),.....,W^b(l+1)], b_i^b 
\end{align*}
The secret-share is performed over a secure channel, and once the share is saved by the two workers, the model owner can go offline.

\subsection{Client}\label{subsec: client}
When the client wants to obtain inference result on input $x$, it connects to workers $A$ and $B$ and secret-shares its private input.  Moreover, it generates a private-public key pair and sends the public key to the aggregator so that when the inference result is ready, before sending it the aggregator encrypts it with the client's public key. When it receives an encrypted label from the aggregator, the client performs a decryption operation using its private key and produces the clear-text label for input $x$. In Section \ref{subsec: aggragator}, we will describe how the aggregator produces the final inference result. Next, we describe how the client secret-shares private input $x$. 

\par \textbf{Secret-Sharing of Client Input:} For simplicity, consider $x_1$ is the first element of the client's input feature vector $x=[x_1, ..., x_d]$. $x_1$ is divided into two parts to be secret-shared with two workers $A$ and $B$, using Equations \ref{eq:share-a} and \ref{eq:share-b}: 
\begin{equation}\label{eq:share-a}
x_1^a=random(-Q,Q)
\end{equation}
\begin{equation}\label{eq:share-b}
 x_1^b=(x_1-x_1^a) \quad \textrm{Mod}\quad Q
\end{equation}
x$_1^a$ and x$_1^b$ are large numbers that fall in the range $[0,Q-1]$, and do not individually reveal the real data. $x_1$ is obtained by combining $x_1^a$ and $x_1^b$ using Equation \ref{eq:combine-shares} below:
\begin{equation}\label{eq:combine-shares}
 x_1=(x_1^a+x_1^b) \quad \textrm{Mod}\quad Q
\end{equation}

Extending the secret-share mechanism from $x_1$ to the whole feature vector $x$ of dimension $d$, the secret-share of input $x$ will be done by extending Equations \ref{eq:share-a} and \ref{eq:share-b} as follows: 
\begin{align*}
   Worker A: x_1^a,x_2^a,....,x_d^a \\
   Worker B: x_1^b,x_2^b,....,x_d^b
\end{align*}
Note that for $i \in [1, d]$ each $x_i^a$ and $x_i^b$ is respectively computed using Equation \ref{eq:share-a} and \ref{eq:share-b}.

\subsection{Workers}\label{subsec: worker}

Next, let us assume we have our network of neurons with total $l$ number of hidden layers, $d$  be the dimension of input $x$ to be secret-shared with worker $A$ and worker $B$. The output results for the $j^{th}$ neuron of the $1^{st}$ hidden layer for worker $A$ and worker $B$ are computed as follows: 
\begin{align} \label{eq:1st_hidden_layer}
Worker A: r^{a,1}_j(x^a,W^a)&=h(\sum_{i=1}^{d} W^a_{ij}(1) x^a_i +b_j)\\
Worker B: r^{b,1}_j(x^b,W^b)&=h(\sum_{i=1}^{d} W^b_{ij}(1) x^b_i +b_j)
\end{align}
Here, $W^a_{ij}(1)$ and $W^b_{ij}(1)$ are the weight vectors from input to the $j^{th}$ neuron of the $1^{th}$ hidden layer.
For the $1^{st}$ hidden layer, we have the output neurons $[r^{a,1}_1,r^{a,1}_2,....,r^{a,1}_{k_1}]$ for worker $A$ and $[r^{b,1}_1,r^{b,1}_2,....,r^{b,1}_{k_1}]$ for worker $B$. Note that  $k_1$ is the total number of neurons in the $1^{st}$ hidden layer.  These output neurons will make the input vector for the $2^{nd}$ hidden layer. Accordingly, the output of the $j^{th}$ neuron of the $2^{nd}$ hidden layer is obtained as:
\begin{align*} \label{eq:2nd hidden_layer}
Worker A: r^{a,2}_j(x^a,W^a)&=h(\sum_{i=1}^{k_1} W^a_{ij}(2) r^{a,1}_i )\\
Worker B: r^{b,2}_j(x^b,W^b)&=h(\sum_{i=1}^{k_1} W^b_{ij}(2) r^{b,1}_i )
\end{align*}

Similarly, the output of the $j^{th}$ neuron of the $l^{th}$ hidden layer is computed as follows:
\begin{align*} %\label{eq:l_hidden_layer}
Worker A: r^{a,l}_j(x^a,W^a)&=h(\sum_{i=1}^{k_{l-1}} W^a_{ij}(l) r^{a,l-1}_i )\\
Worker B: r^{b,l}_j(x^b,W^b)&=h(\sum_{i=1}^{k_{l-1}} W^b_{ij}(l) r^{b,l-1}_i )
\end{align*}
Here, $W^a_{ij}(l)$ and $W^b_{ij}(l)$ are the weight vectors from the $(l-1)^{th}$ hidden layer to the $j^{th}$ neuron of the $l^{th}$ hidden layer.

Each worker uses the secret-shared private input of the client and private model parameters of each model owner to produce intermediate results, which, when combined, produce the true inference output. Finally, the intermediate results for worker $A$ and worker $B$ for the $j^{th}$ output neuron are computed as:
\begin{align}%\label{eq:outNN}
Worker A: y^a_j(x^a,W^a)&=h(\sum_{i=1}^{k_{l}} W^a_{ij}(l+1) r^{a,l}_i )\\
Worker B: y^b_j(x^b,W^b)&=h(\sum_{i=1}^{k_{l}} W^b_{ij}(l+1) r^{b,l}_i )
\end{align}

The coefficients $W^a_{ij}(l+1)$ and $W^b_{ij}(l+1)$ are the weight vectors from the final hidden layer $l$ to the $j^{th}$ neuron of the output layer for worker $A$ and worker $B$, respectively.
%The $\xi^a$ =$(\xi^a_0,\xi^a_1,.......\xi^a_k)^T$ and $\xi^a$ =$(\xi^b_0,\xi^b_1,.......\xi^b_k)^T$ are a vector of non-linear parametric basic functions for worker 'A' and 'B' respectively  The $j^{th}$ basis function for worker "A" and worker "B" can be represented as:
%\begin{align} \label{eq: worker_basisFn}
%Worker A: \xi^a_j(x)&=h(\sum_{i=1}^{n} W^a_{ij}^{(1)} x^a_i +b^a_j)\\
%Worker B: \xi^b_j(x)&=h(\sum_{i=1}^{n} W^b_{ij}^{(1)} x^b_i +b^b_j)
%\end{align}

%The output result of $r^{th}$ neuron of output layer $l$ of model owner $P_i$ is represented as, 
%\begin{align}\label{eq:outworker}
%Worker A: y^a_{lr}(x^a,W^a)&=\sum_{j=0}^{k} W^a_{rj}(l) h(\sum_{i=0}^{d} W^a_{ij}(1) x^a_i +b^a_j)\\
%Worker A: y^b_{lr}(x^b,W^b)&=\sum_{j=0}^{k} W^b_{rj}(l) h(\sum_{i=0}^{n} W^b_{ij}(1) x^b_i +b^b_j)
%\end{align}
Finally,  the resultant intermediate vector for the output layer of model owner $P_i$ for $o$ number of neurons is $y_{P_i}^a=[y^a_{P_i,1}, y^a_{P_i,2}, ...., y^a_{P_i,o}]$ for worker $A$ and $y_{P_i}^b=[y^b_{P_i,1},y^b_{P_i,2}, ...., y^b_{P_i,o}]$ for worker $B$, respectively. These final vectors $y_{P_i}^a$ and $y_{P_i}^b$ are sent to the aggregator via a secure communication channel.

\subsection{Aggregator}\label{subsec: aggragator}
\textbf{Reconstruction of Inference Results:} When it receives intermediate results $y_a$ and $y_b$ from the workers, the aggregator first combines $y_a$ and $y_b$ to obtain the true inference result $y$.  Across each $y$ that corresponds to a model owner, the aggregator then performs majority vote-based aggregation of inference results. The aggregator then adds random noise sampled from the Laplace distribution to perturb the inference output (and hence deter membership inference style attacks). Lastly, the aggregator encrypts the inference output with client's public key before sending it to the client. 

\par In general, for each model owner $P_i$, the aggregator receives $y_{P_i}^a=[y^a_{P_i,1}, y^a_{P_i,2},.., y^a_{P_i,o}]$ and $y_{P_i}^b=[y^b_{P_i,1}, y^b_{P_i,2}, .., y^b_{P_i,o}]$  from worker A and B respectively. To get the true inference value from the intermediate results, the aggregator combines them as follows:

\begin{align}\label{eq:combine-inference}
    y_{P_i} = (y_{P_i}^a+y_{P_i}^b )\quad Mod \quad (Q)
\end{align}
As a result, the constructed inference result for model owner $P_i$ is $y_{P_i}=[y_{P_i,1}, y_{P_i,2}, ...., y_{P_i,o}]$ for $o$ number of output neurons. Hence, the output for $m$ model owners is represented as a vector of final inference results: $[y_{P_1}, y_{P_2},....,y_{P_m}]$. 

\textbf{Noisy Aggregation of Inference Results:} The aggregator aggregates all the inference results for $m$ model owners as follows: 
\begin{align}\label{eq:add-noise}
    y_{noise}=\Phi_{i=0}^{m} y_{P_i} + Lap(\frac{s}{\epsilon})
\end{align}
In Equation \ref{eq:add-noise} above, $y_{noise}$ is the noisy aggregated output for $m$ number of model owners, and $\Phi$ is the aggregation function (e.g., majority vote, confidence-weighted sum of inference probabilities). Here, $y_{noise}=[y_1,y_2,...,y_o]$ is an $o$-dimensional vector with a total of $o$ number of noisy aggregated output inferences. The final label is produced using the $argmax$ operation as $y=argmax([y_1, y_2, ..., y_o])$, where $y$ is the output label for input vector $x$. The aggregator encrypts this output with the client's public key and sends it to the client over a secure channel.
\section{Evaluation}\label{sec: eval}
In this section, we valuate \sysname by answering the following research questions:

\textbf{\textit{RQ1}:}. What is the accuracy/privacy trade-off for \sysname and how does it compare with prior work?

\textbf{\textit{RQ2}:} What is the impact of the number of model owners on inference accuracy and privacy?

\textbf{\textit{RQ3}:} What is the impact of differentially-private aggregation in reducing membership inference attack by a semi-honest client?

\textbf{\textit{RQ4}:} What performance overhead is incurred by \sysname overall, per-sample, and per-model?

\subsection{Datasets, Model Architectures, and Setup}
We use four datasets that focus on: handwritten digit recognition (MNIST~\cite{MNIST}), clothing image classification (Fashion-MNIST~\cite{Fashion-MNIST}), breast  
cancer detection (IDC~\cite{IDC}), and  in-ICU length-of-stay prediction for patients (MIMIC~\cite{MIMIC}). We chose these datasets because they were used as benchmarks in related work~\cite{PATE17,SecureML17,Chameleon, CryptoNets16} and also are relevant for privacy-sensitive domains (e.g., IDC and MIMIC are medical datasets). Next, we first briefly describe each dataset and then discuss model architectures and training setup.

\textbf{MNIST} is a collection of 70K gray-scale images of handwritten digits with training set of 60K images and a test set of 10K images. Each sample has a width-height dimension of 28x28 pixels of handwritten digits 0 to 9, which make up the 10 classes. We divide the 60K training samples into multiple chunks to train multiple models. For instance, for 50 model owners, we divide the training set into 50 disjoint datasets so that each model owner has 1.2K samples to train their model on. We use 500 samples from the test set for the evaluation of \sysname.

\textbf{Fashion-MNIST} is a dataset from Zalando's article images with 60K training set and 10K test set. Each example is a 28x28 gray-scale image, associated with a label from 10 classes. The labels are for T-shirt/top, trouser, dresses, etc. For the highest number of model owners of 40, we divide the training set into 40 disjoint datasets so each model owner has 1.5K samples to train their model. We use 500 samples from the test set to evaluate \sysname.

\textbf{IDC} is the Invasive Ductal Carcinoma (IDC) dataset with 277,524 patches of size 50×50 extracted from 162 whole mount slide images of breast cancer specimens scanned at 40x. Out of the 277,524 samples, 198,738 test is negative (benign) and  78,786 test positive with IDC. Using under-sampling to balance positive and negative samples, we use 277,024 samples for training a binary classifier and 500 samples as testing samples for \sysname. 

\textbf{MIMIC} is the Medical Information Mart for Intensive Care (MIMIC) dataset with 60K samples with details such as demographics, vital signs, laboratory tests, medications and personal information about patients. The classifier predicts the length of stay (LOS) of a patient in the hospital. We use a multi-class classifier where LOS is divided into four classes: class 0 (0-4 days), class 1 (4-8 days), class 2 (8-12 days), and  class 3 (12 or more days). Our training dataset contains 58, 386 samples and test set has 590 samples. 

\textbf{Model Architectures and Setup:}
For all the four datasets, we train a FFNN model. For MNIST and Fashion-MNIST, the input layer is $28\times28$ and the output layer has $10$ neurons, with two hidden layers with {\em ReLU} activation. The output layer is linear and we use {\em Mean Square Error}  loss instead of the negative log-likelihood function. Since log-likelihood is performed using {\em softmax} function, it requires computing  logarithm and exponential functions which are not practical for \sysname because we convert all real number parameters to integers. Since our computations are over a finite integer field, we convert all the floating point tensors into fixed precision tensors with a rounding at the second decimal digit (e.g., .456 to 45). For MIMIC, we use one hidden layer with $30$ input neurons and $4$ output neurons. For IDC, we use the same FFNN model where the number of input neurons is $7500$ and the number of output neurons is 2. Model architecture details are described in the Appendix (Section \ref{subsec: model-architectures}). The learning rate is 0.001 across the four datasets. The number of epochs for MNIST and Fashion-MNIST is $100$, while we used $80$ epochs for MIMIC and $200$ epochs for IDC.  Finally, for each dataset, each model owner secret-shares their respective model parameters with workers A and B.  

\subsection{Accuracy/Privacy Trade-Off}\label{subsec: acc-priv-trade-off}
\begin{figure*}[htb!]
\centering
\begin{minipage}[t]{0.24\textwidth}
%  \centering
 \includegraphics[scale =.29]{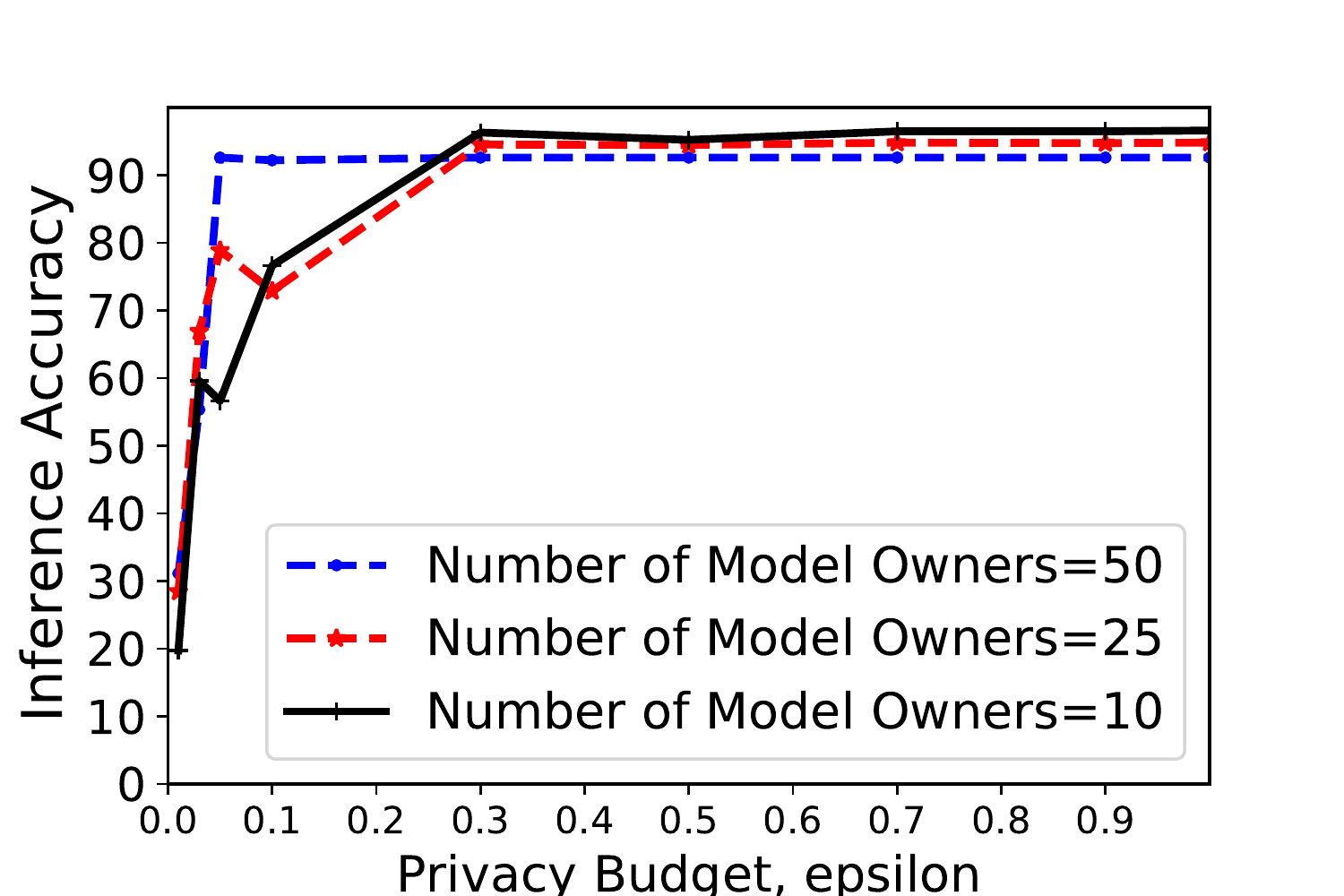}
  \caption{\scriptsize MNIST: Accuracy/privacy trade-off for $\epsilon \in [0.01,1]$ and $m \in \{10, 25, 50\}$.}
\label{fig:MNIST}
\end{minipage}
\hspace{0.03cm}
\begin{minipage}[t]{0.24\textwidth}
%  \centering
 \includegraphics[scale=.29]{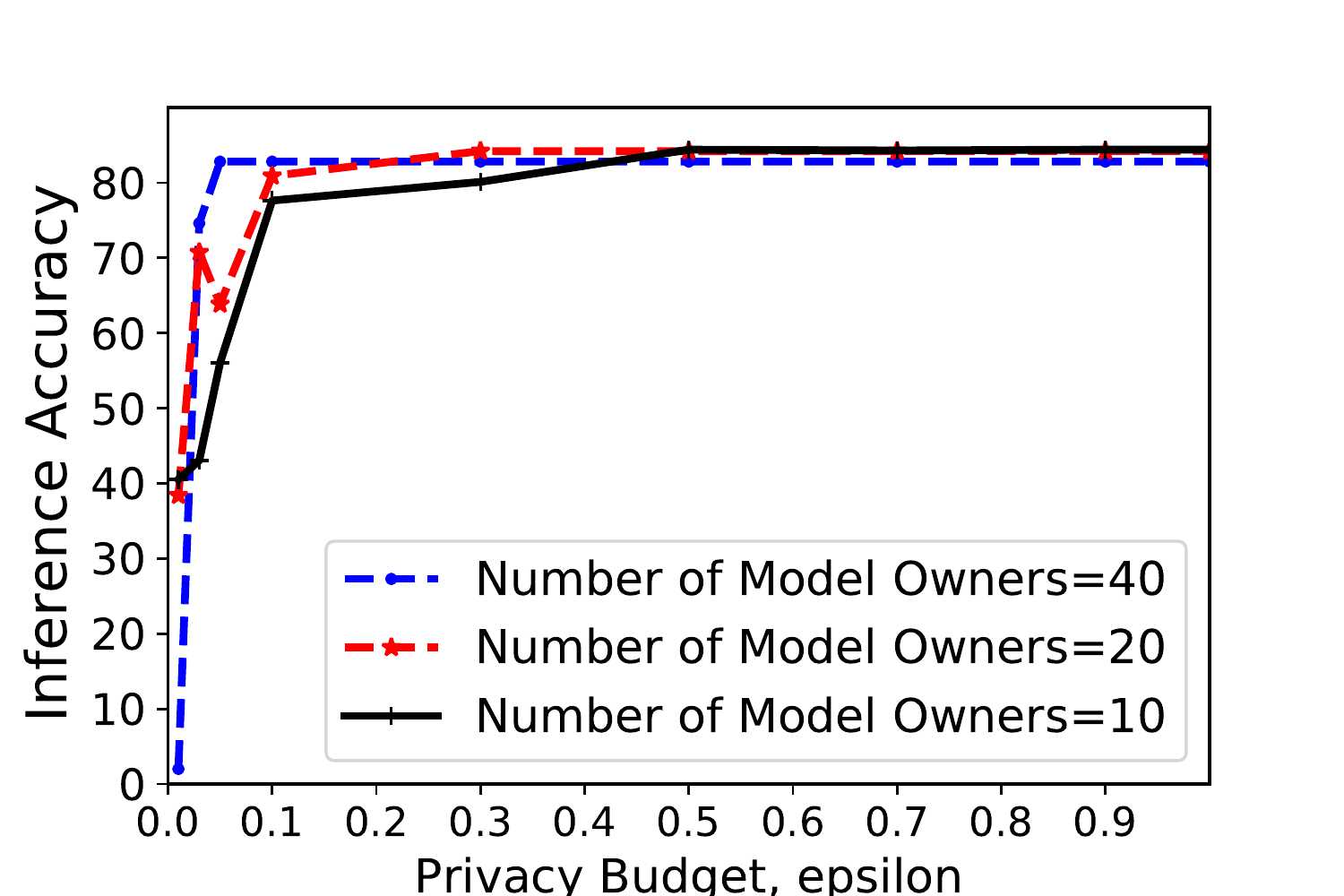}
 \caption{\scriptsize Fashion-MNIST: Accuracy/privacy trade-off for $\epsilon \in [0.01,1]$ and $m \in \{10, 20, 40\}$.}
\label{fig:Fashion-MNIST}
\end{minipage}
\begin{minipage}[t]{0.24\textwidth}
%  \centering
 \includegraphics[scale=.29]{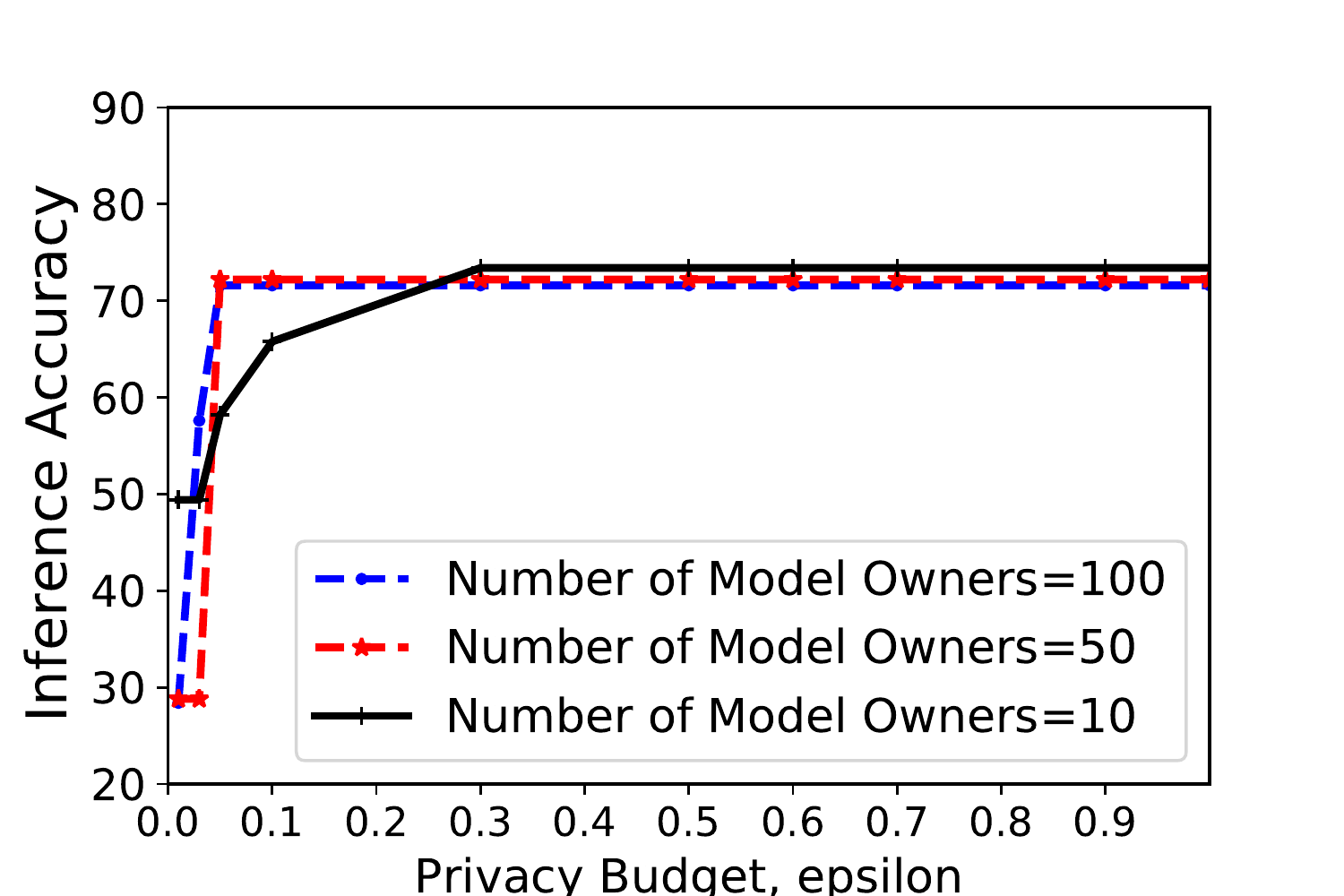}
 \caption{\scriptsize IDC: Accuracy/privacy trade-off for $\epsilon \in [0.01,1]$ and $m \in \{10, 50, 100\}$.}
\label{fig:IDC}
\end{minipage}
\hspace{0.03cm}
\begin{minipage}[t]{0.24\textwidth}
%  \centering
 \includegraphics[scale=.29]{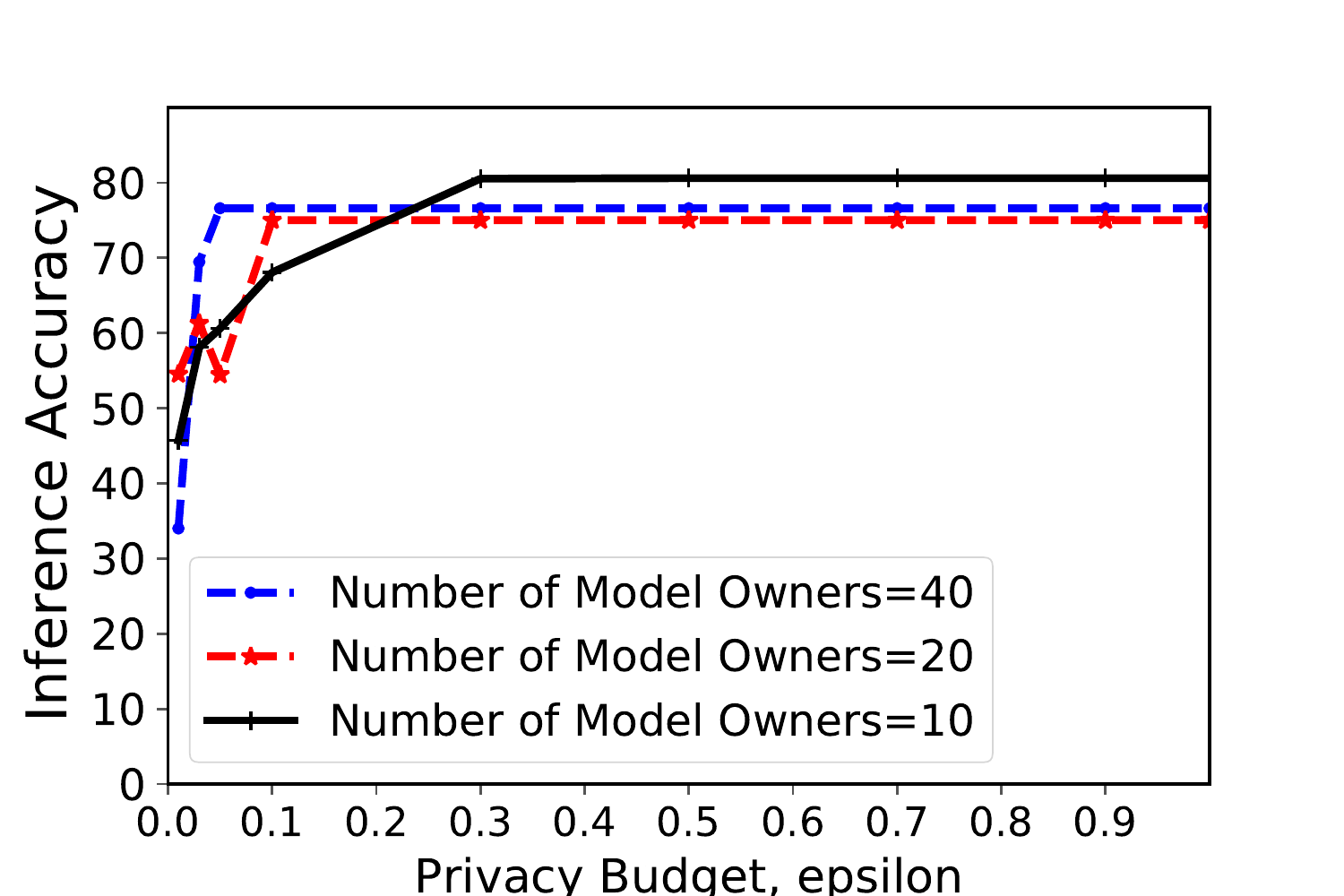}
 \caption{\scriptsize MIMIC: Accuracy/privacy trade-off for $\epsilon \in [0.01,1]$ and $m \in \{10, 20, 40\}$.}
\label{fig:Hospital}
\end{minipage}
\end{figure*}

We first analyze the inference accuracy of \sysname with respect to privacy budget $\epsilon$. Smaller $\epsilon$ values imply stronger privacy guarantees (e.g., against membership inference) and vice-versa. The ideal/optimal case is when we achieve high inference accuracy while providing strong privacy (smallest $\epsilon$ value), but striking the sweet spot that fulfills both is generally non-trivial for the trade-off depends on training data, number of model owners, and aggregation scheme used to decide the final inference result. In Figures \ref{fig:MNIST}--\ref{fig:Hospital}, $x$-axis shows privacy budget $\epsilon$ and $y$-axis is inference accuracy of \sysname. Across the four datasets, the smallest number of model owners is $m = 10$, while the maximum is in the range $40-100$ (e.g., $40$ for Fashion-MNIST, $50$ for MNIST, and $100$ for IDC), depending on the size of the training set and inference accuracy.

\textbf{On MNIST and Fashion-MNIST Benchmarks:}
Figure $\ref{fig:MNIST}$ shows inference accuracy with respect to privacy budget $\epsilon$ (in the range [10$^{-2}$, 1]) for \sysname. Note that we have divided the $60$K MNIST training set into range of 10, 25, and 50 disjoint datasets with $6K$, $2.4K$, and $1.2K$ per-model owner samples, respectively. The client has $500$ input samples as the inference set. For MNIST, when number of model owners $m=50$, the inference accuracy is $55.53\%$ for $\epsilon=0.03$. Compared to the non-private $92.4\%$ inference accuracy for MNIST, we observe that \sysname incurs significant accuracy loss for the privacy gain it provides in exchange. Interestingly, for $\epsilon=0.05$, the inference accuracy for \sysname is $92.6\%$ (i.e, no accuracy loss), and the inference accuracy remains fairly stable for $\epsilon \in [0.5, 1]$ (see Figure \ref{fig:MNIST}). This observation is consistent with prior work on ensemble of differentially private teacher models by Papernot et. al.~\cite{PATE17, ScalablePATE18}. 
The best trade-off is with $\epsilon=0.05$ and inference accuracy=92.6\% when $m = 50$.

%ACCURACY/PRIVACY TRADE-OFF

 Figure $\ref{fig:Fashion-MNIST}$ shows accuracy/privacy trade-off analysis for Fashion-MNIST for 10, 20, and 40 model owners. For $m = 40$, the inference accuracy is 82.8\% with $\epsilon=0.1$. For  $\epsilon < 0.1$, the inference accuracy drops dramatically. For example, when $\epsilon=0.05$, inference accuracy is $74.6\%$ which is smaller than our best case accuracy on Fashion-MNIST of $82.8\%$ when $m=40$. Note that non-private inference accuracy is $82.8\%$, which, similar to MNIST, suggests that there is no accuracy loss. 
The best trade-off is obtained with $\epsilon=0.1$ and inference accuracy = 82.8\% for $m=40$.

\textbf{On IDC and MIMIC:} Figure $\ref{fig:IDC}$ shows accuracy/privacy trade-off analysis for IDC. When $m=100$, inference accuracy is 71.6\% for $\epsilon$=0.05, which is the optimal case in terms of accuracy/privacy trade-off. Hence, \sysname offers an acceptable inference accuracy for IDC with smaller privacy budget, which offers a strong privacy guarantee. For $\epsilon < 0.05$, the inference accuracy is lower, for example, for $\epsilon=0.03$, inference accuracy is 71.6\% with $m=100$, which suggests that the acceptable noise range for IDC is $\epsilon \in [0.05, 1]$. Figure \ref{fig:Hospital} shows accuracy/privacy trade-off for MIMIC. The best case inference accuracy is 76.6\% with $\epsilon =0.05$ for $m=40$. For $\epsilon <0.05$, the inference accuracy is much lower (e.g., $34\%$ for $\epsilon =0.01$).

\textbf{Comparison with Closely Related Work:} Papernot et. al.\cite{PATE17} show with a CNN model on MNIST, their non-private model is 99.18\% accurate while the private ensemble of 250 teacher models is 93.18\% accurate. In our case, the non-private FFNN model on MNIST is 96.6\% accurate, while its private counterpart of 50 model owners is 92.6\% with $\epsilon = 0.05$. We note that initial differences in the non-private accuracy of the models could be attributed to the different in model architectures, i.e., CNN (theirs) vs. FFNN (ours).

$\bigstar$ \textbf{Take-away:} With respect to \textbf{\textit{RQ1}}, our analysis of the accuracy/privacy trade-off suggests that \sysname provides acceptable privacy guarantees with very minimal trade-off on inference accuracy, showing its  practical viability in a multi-party setting.

\subsection{Number of Model Owners vs. Accuracy and Privacy}\label{subsec: model-ownwers-vs-acc-priv}
As we indicated in Section \ref{subsec: model-owner}, intuitively, a higher value for number of model owners  $m$ would potentially entail better collective inference accuracy. However, this depends on the accuracy of each model and the privacy budget. With that background, here, we analyze the impact of $m$ on inference accuracy and privacy budget.

\textbf{On MNIST:} 
From  Figure \ref{fig:MNIST}, as $m$ increases from 10 to 25 and then 50, we notice a roughly consistent trend, especially for $\epsilon > 0.03$. With progressive increase in $\epsilon$ from  $\epsilon = 0.05$ to $\epsilon = 1$, inference accuracy seems to be inversely proportional to increase in $m$. Keeping the same trend for $\epsilon \in [0.01,1]$, the highest inference accuracy values are achieved for the lowest number of model owners ($m =10$), and the lowest inference accuracy values correspond to the highest number of model owners ($m = 50$). In particular, for $m = 50$, the inference accuracy is 92.4\% (with $\epsilon=0.05$). On the other hand, inference accuracy jumps to 94.52\% (with $\epsilon=0.3$) and then  96.56\% (with $\epsilon=0.5$) for $m=25$ and $m=10$, respectively.

Overall, from these observations we synthesize that higher number of model owners (e.g., $m=50$) result in lower inference accuracy as opposed to lower number of model owners (e.g., $m =25$, $m = 10$). It is, however, noteworthy that higher inference accuracy for lower number of model owners offer relatively lower privacy guarantees. For instance, with $m =25$ and $m = 10$, inference accuracy of 94.52\% and 96.56\% are achieved with $\epsilon = 0.3$ and $\epsilon = 0.5$, respectively.

\textbf{Fashion-MNIST, IDC, and MIMIC:}
We observe similar results for Fashion-MNIST and IDC from Figure $\ref{fig:Fashion-MNIST}$ and Figure $\ref{fig:IDC}$. For Fashion-MNIST, the best case of inference accuracy is 82.8\% for $\epsilon=0.03$, 84.2\% for $\epsilon=0.3$, and 84.39\% for $\epsilon=0.5$, respectively, for 40, 20, and 10 model owners. On the other hand, for IDC, the best case of inference accuracy is 71.6\% for $\epsilon=0.05$, 72.2\% for $\epsilon=0.05$, and 73.4\% for $\epsilon=0.3$, respectively, for 100, 50, and 10 model owners. As a result, for Fashion-MNIST and IDC dataset, we observe that \sysname achieves better inference accuracy for smaller number of model owners, but with a caveat of relatively lower privacy guarantee.
For MIMIC, from Figure $\ref{fig:Hospital}$ we observe that for 40 model owners, the best case for inference accuracy is 76.6\% with $\epsilon=0.05$, while for 10 model owners, the best case for inference accuracy is 80.53\% with $\epsilon=0.3$ which also supports our claims.

$\bigstar$ \textbf{Take-away:}
With respect to \textbf{\textit{RQ2}}, we note that if the number of model owners $m$ is higher, it allows a larger noise level (i.e., offers stronger privacy guarantee), while for smaller $m$ values the inference accuracy is more sensitive to noise. Our findings here are inline with PATE \cite{PATE17} with respect to larger number of models resulting in lower privacy budget on MNIST.

\subsection{Utility of Differential Privacy Against Membership Inference Attack}\label{subsec: MIA-results}
We recall that one of the goals of \sysname is to deter a semi-honest client who may mount membership inference attack (MIA). To that end, we perform DP-aggregation of inference results before we release each inference result to the client. The goal of MIA is to identify (with some confidence) whether a given sample belongs to a training set of a target model~\cite{MIA}. To examine the utility of DP in limiting MIA, we measure the accuracy, precision and recall of MIA for both noiseless and noisy aggregation of inference results. {\em Accuracy} measures the percentage of examples that are correctly predicted to be members of the target model’s training dataset. {\em Precision} measures the proportion of true membership inference with respect to all reported attacks, while {\em recall} measures the coverage of the attack, as the fraction of the training records that the attacker can correctly deduce as members of the training set. For the purpose of this evaluation, we reproduce the MIA by Shokri et. al. \cite{MIA}, and we use the same size of member and non-member data to maximize the inference uncertainty, with baseline accuracy of $50\%$. We use Fashion-MNIST and MIMIC to examine the effect of differential privacy on MIA.
\begin{figure}[b!]
\centering
\begin{minipage}[t]{0.48\columnwidth}
  \centering
 \includegraphics[scale=.3]{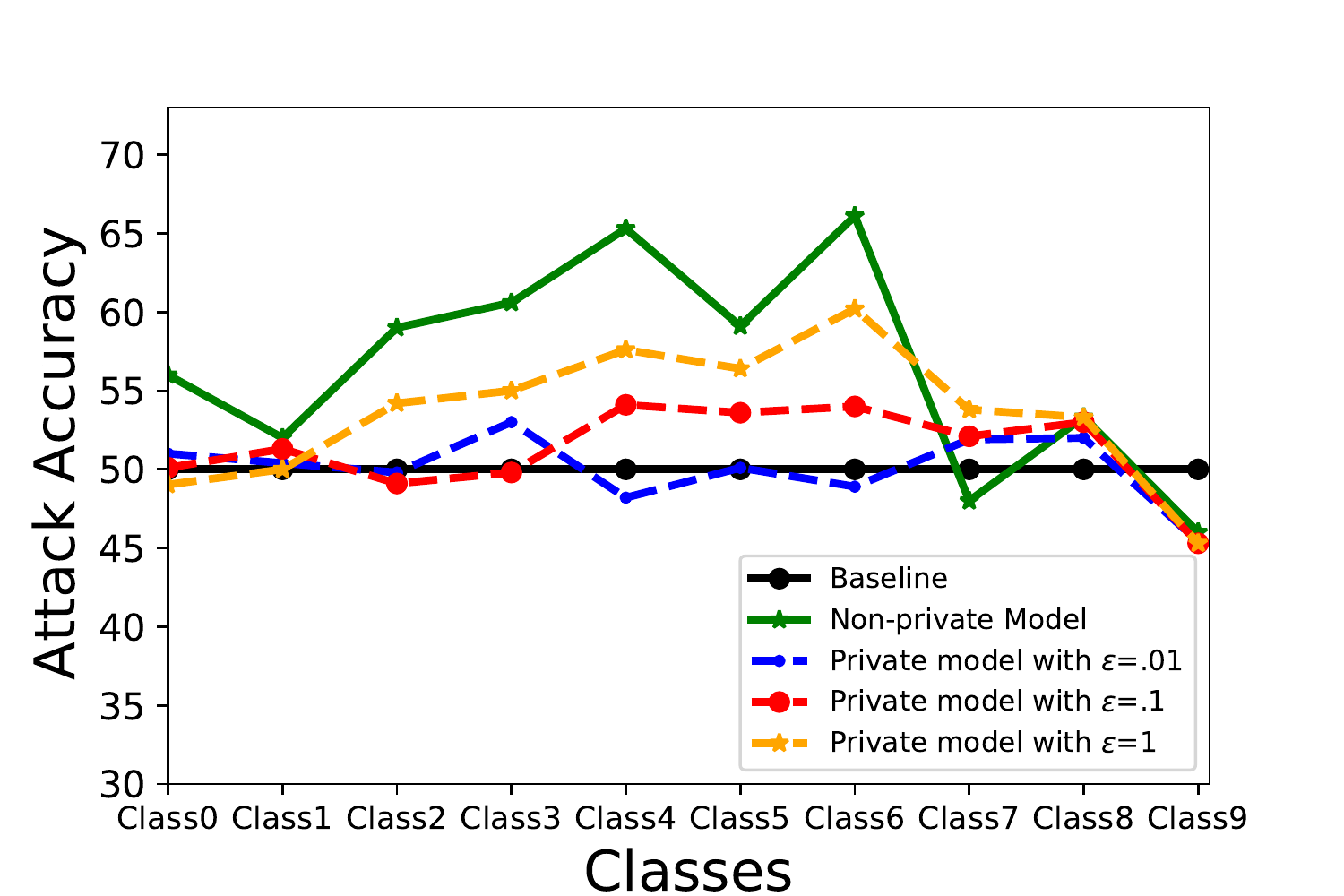}
  \caption{\scriptsize Membership inference attack accuracy comparison for non-private and $\epsilon$-DP model on Fashion-MNIST.}
 \label{fig:MIA_FMNIST}
\end{minipage}
% \hspace{0.01cm}
\begin{minipage}[t]{0.48\columnwidth}
  \centering
 \includegraphics[scale=.3]{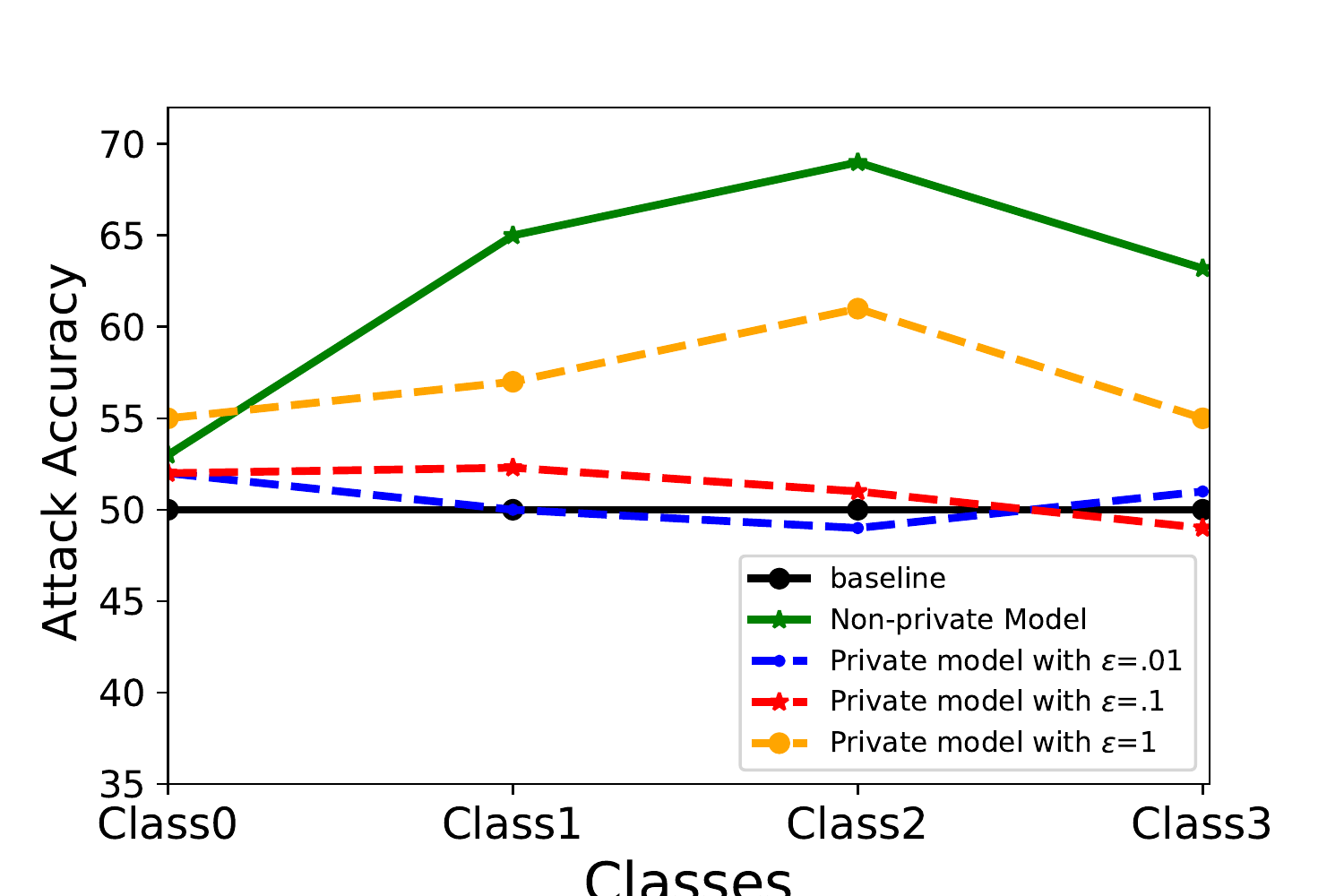}
  \caption{\scriptsize Membership inference attack accuracy comparison for non-private and $\epsilon$-DP model on MIMIC.}
\label{fig:MIA_MIMIC}
\end{minipage}
\end{figure}
\vspace{-.2em}

\textbf{MIA Against Fashion-MNIST Model:}
We use 5,000 samples to train the target model. There are $20$ shadow models and their training size is set to $5, 000$.  Training dataset of shadow models are disjoint with the training dataset of the target model. The shadow models are expected to mimic the behavior of the target model as the target model and the shadow models are all trained on data coming from the same population.  As a test dataset, we use the samples with equal number of members and non-members. 

Figure \ref{fig:MIA_FMNIST} shows per-class accuracy of MIA for both private and non-private model on Fashion-MNIST.  Without differentially private prediction, the maximum, minimum, and average MIA accuracy, respectively, is $69.3\%$ (for class-4), $46\%$ (for class-9), and $57.13\%$ (for all 10 classes). On the other hand, when we add Laplacian noise with $\epsilon=0.1$ post-aggregation on each inference result, average MIA accuracy is $51.17\%$ with maximum of $54.1\%$ for class-4 and minimum accuracy of $45.3\%$ for class-9. The attack accuracy degradation (by 5.96\% to be exact) implies that due to the presence of Laplacian noise, the model owner's training data are less vulnerable to MIA. We also observe that, adding more noise gradually decreases the MIA accuracy, thus mitigating training data exposure. 

The average precision value for all class without DP is $54.22\%$, suggesting that for the non-private model, $54.22\%$ of the images that were inferred as members by the attacker are true members. On the other hand, average recall score for the non-private model for all classes is $79.02\%$, implying that the attacker can correctly infer an averages of $79.02\%$ of the training images as members. On the other hand, precision and recall with privacy budget = $0.1$ are $51.2\%$ and $46.24\%$, respectively. The lower precision (by 3.02\%) and lower recall (by 32.78\%) for the private model suggest that  $\epsilon$-DP mitigates the membership inference attack.

\textbf{MIA Against MIMIC Model:}
We use $5, 000$ samples for both the target model and shadow model, with $20$ shadow models. We use the same number of members and non-members for evaluation. Figure \ref{fig:MIA_MIMIC} shows per-class accuracy of MIA for both private and non-private model on MIMIC. In a non-private setting, the maximum MIA accuracy is $69\%$ for class-2 and minimum MIA accuracy is $53\%$ for class-0. The average MIA accuracy is $61.18\%$ for all four classes, suggesting $61.18\%$ examples of the target model’s training data are correctly predicted to be members. With the privacy budget $\epsilon=0.1$, the maximum MIA accuracy is $52.3\%$ for class-1 and minimum MIA accuracy is $49\%$ for class-3. For these settings, the average accuracy of MIA is $52.1\%$ across all four classes. The average accuracy degradation is $9.08\%$, which suggests that $\epsilon$-DP mitigates MIA by $9.08\%$. Looking at precision and recall for the non-private model, we obtain $56.47\%$ and $96.2\%$, respectively. On the other hand, precision and recall with privacy budget = $0.1$ are $50.95\%$ and $51.48\%$, respectively. The lower precision (by 5.52\%) and lower recall (by 44.72\%) for the private model suggest that  $\epsilon$-DP mitigates the attacker's pursuit of members' records. 

\textbf{Comparison with Closely Related Work:} While PATE \cite{PATE17} discusses intuitive guarantee against MIA, they did not provide quantitative measurement of MIA based on noise scale. In \cite{MIA_DP}, the effect of DP-based training against MIA is examined. On MNIST, with $\epsilon=2$ and $\epsilon=1$, they show that DP can eventually mitigate MIA accuracy and bring it close to baseline. Unlike \cite{MIA_DP}, in \sysname we use $\epsilon$-DP after aggregating the inference results of model owners, with sizable MIA mitigation within our noise boundary.

$\bigstar$ \textbf{Take-away:} With respect to \textbf {\textit{RQ3}}, differentially private aggregation limits membership inference attack in a way that both the inference about members and the coverage of attack are minimized. 

\subsection{Performance Overhead Analysis}\label{subsec: preformance-overhead}

\begin{figure*}[t!]
\centering
\begin{minipage}[t]{0.45\textwidth}
\centering
 \includegraphics[scale =.46]{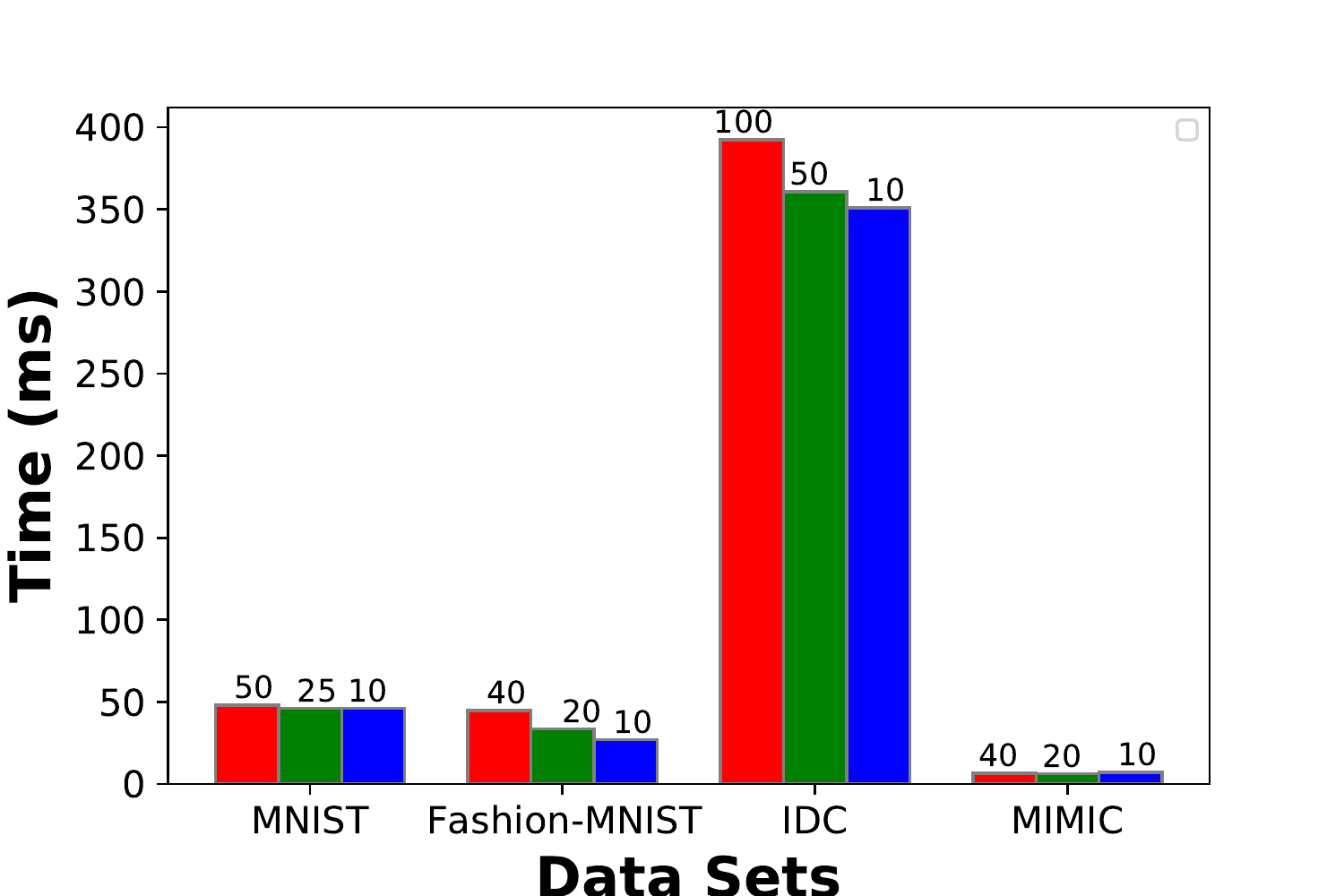}
  \caption{\scriptsize Per-model secret-sharing overhead in milli-seconds.}
\label{fig:Performance_Owner}
\end{minipage}
\hspace{0.08cm}
% \begin{minipage}[t]{0.45\textwidth}
%  \centering
%  \includegraphics[scale=.37]{figs/bar2.pdf}
%  \caption{\scriptsize Per-sample noisy aggregation overhead in micro-seconds for MNIST, Fashion-MNIST, IDC, and MIMIC.}
% \label{fig:Performance_AGG}
% \end{minipage}
\begin{minipage}[t]{0.45\textwidth}
 \centering
 \includegraphics[scale=.46]{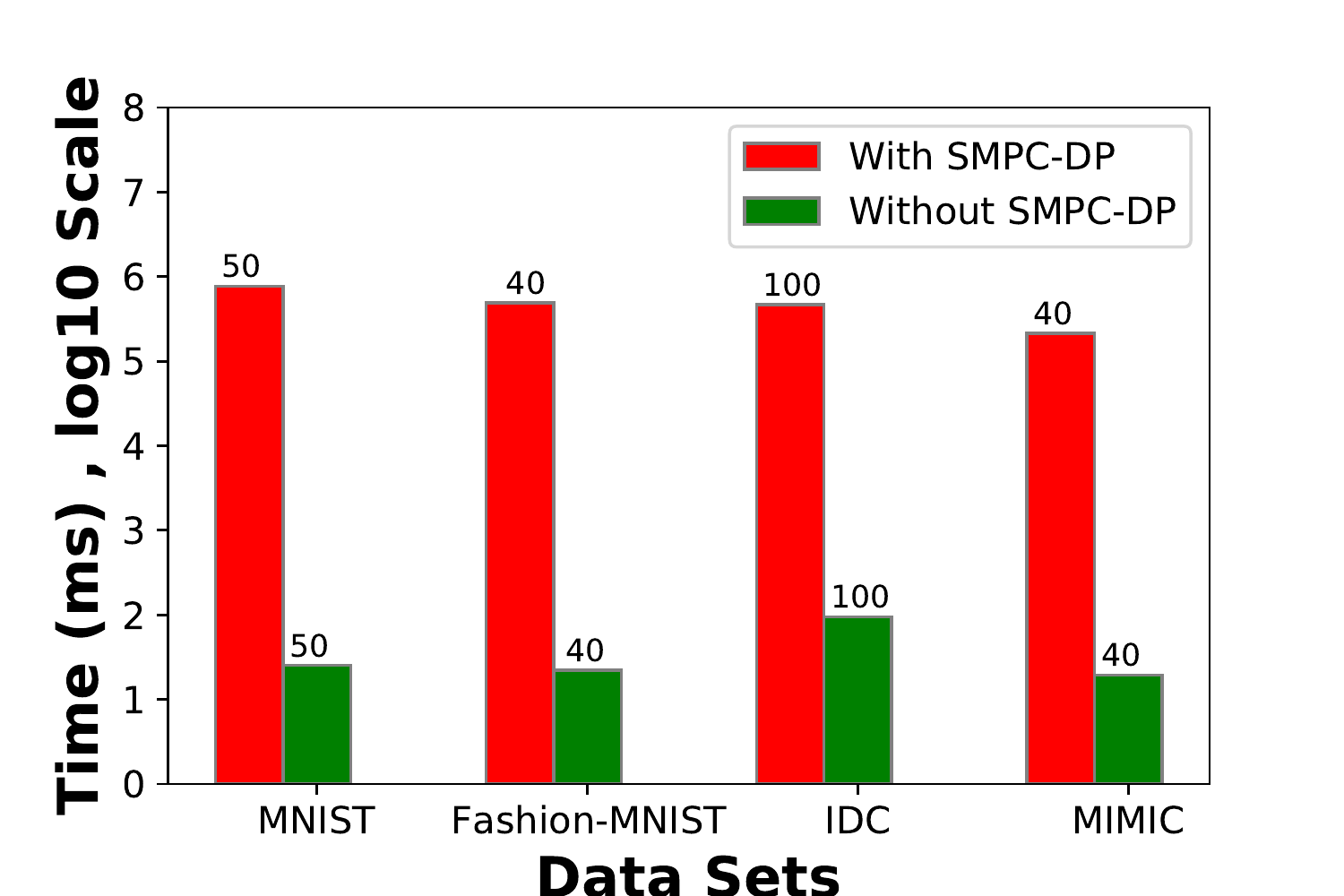}
 \caption{\scriptsize Overall per-sample inference overhead in milli-seconds with SMPC-DP (\sysname) and without SMPC-DP (non-private).}
\label{fig:Performance_overall}
\end{minipage}
\end{figure*}

We evaluate the overall, per-sample, and per-model performance overhead incurred by \sysname. In particular, we measure time taken by a) a model owner to secret-share their model parameters with the workers and b) per-sample overhead for inference. 

\textbf{Hardware Specification:} 
To run our experiments, we used Google Colab virtual machine which provides 2 Intel(R) Xeon(R) model 79, family 6 CPUs with 2.2GHz each, with 12.72GB RAM and 107.77GB disk space. To speed-up computations, we used Python's multi-threading features to efficiently utilize the two CPUs at a time, by dividing the task between them and get output inferences from two model owners in one cycle. 

\textbf{Per-model Overhead for Secret-Sharing:} Figure \ref{fig:Performance_Owner} shows the duration of secret sharing of model parameter across the four datasets, and for different number of model owners. This performance is influenced by the number of model parameters ($W$ and $b$). The $x$-axis shows the different datasets and the $y$-axis shows the time needed, in milliseconds, to share a single model owner's model parameters with the two workers. Overall, the per-model secret-sharing overhead is insignificant. Comparatively, IDC seems to incur more delay (up to $392.4$ms) of model parameter secret sharing because it produces vast model parameters compared to others.  The other data sets need much less time in the range 8ms-48ms to share their model parameters, which is insignificant.

\textbf{Per-sample Overhead for Inference:}
In Figure \ref{fig:Performance_overall}, the $y$-axis shows overall duration (in $log_{10}$ scale) to produce inference result for an input with SMPC-DP (\sysname) and without SMPC-DP (non-private). This performance is affected by number of model owners $m$, number of classes, and feature vector dimension of the input. From Figure \ref{fig:Performance_overall}, across the board \sysname incurs a significant overhead to produce the inference result. For example, on MNIST, \sysname requires 772.56s to produce an inference result for one input (with $m=50$), while producing non-private prediction in just 0.025s. While the overhead seems significant on the surface, given the commodity hardware we used, which can be accelerated with parallelization and GPUs, we do not consider the overhead to be a deal-breaker for practical deployability of \sysname.

\textbf{Comparison with Closely Related Work:}
{\it SecureNN}~\cite{SecureNN} proposes a DNN-centric protocol for secure training and prediction of 3- and 4-party setup. It was ran on Amazon EC2 c4.8x large instances in a LAN setting and WAN setting with average bandwidth of 625MB/s and 40MB/s, respectively. For per-sample inference on MNIST, 3-party and 4-party SecureNN require 0.34sec and 0.23sec, respectively. On the other hand, with 50-party setup ($m=50$), \sysname requires 772.56sec to compute inference result for an image. Note that for $m=1$, \sysname takes $1.5$sec to compute a per-sample prediction, which is comparable to SecureNN given the difference in hardware specifications and latency setup. 

{\it Chameleon}~\cite{Chameleon} is evaluated based on a 5-layer CNN on MNIST with inference label as output. Experiments were ran on Intel Xeon ES-1620 CPU, 3.5 GHz with 16 GB of RAM. Per-sample classification took 2.7sec with testing accuracy of 99\%. In \sysname, $m$ is orders of magnitude larger (e.g., $m = 50$, $m = 100$, depending on the dataset) and the inference is collaborative.

{\it CryptoNets}~\cite{CryptoNets16} is a cloud-based platform where the server can provide encrypted prediction to client's computing on their encrypted data so both parties' data can be secure. CryptoNets was ran on a similar hardware as \cite{SecureML17} and per-sample classification took 297.5s with 99\% test accuracy.

$\bigstar$ \textbf{Take-away:} With regards to \textbf{\textit{RQ4}}, secret-sharing model parameters and aggregating inference vectors incur negligible overhead. We also note that inference overhead in \sysname is fairly significant compared to its non-private counterpart. We, therefore, see room for optimizations through hardware acceleration, parallelization, and efficiency advances in SMPC. 

\subsection{Discussion on Limitations}
\textbf{Scalability:} \sysname's effectiveness depends on: number of model owners, and complexity of models and inputs. Other factors kept constant, increasing number of model owners slows down the inference process, because more models means more load on the workers. Model complexity/architecture plays an important role in system performance. Model details such as number of hidden of layers, activation functions, and pooling operations determine the effectiveness of \sysname. The complexity of feature vectors of inputs also determines the speed of computing predictions and the complexity of the learned parameters. For instance, images of 28x28 pixels for MNIST result in a model far less lightweight than images of 50x50 pixels for IDC. While our evaluations are encouraging as to real-life deployability of \sysname, its scalability with more complex model architectures is worth exploring as future work. In addition, as the number of models increase, at some point, the performance load on the workers will be too much to bear, which demands performance enhancement alternatives such as hardware acceleration, multi-threading, and introducing more workers (albeit more complexity).  

\textbf{Performance Overhead Measurements:} Due to our limitations of hardware resources, we measured performance overhead of \sysname on a single machine. As a result, some measurements (e.g., secret-sharing delay, sending intermediate results from workers to the aggregator) may not fully capture an actual distributed setup. As in prior work (e.g., \cite{SecureNN}), production performance overhead measures need to be done in distributed setup to measure network latency for secret-sharing model parameters, computing aggregated inference results, and sending them to the client.

\section{Related Work}\label{sec: related}
We discuss related work in the context of \sysname. For detailed survey, we refer the reader to \cite{survey_ML1} and \cite{SP-SoK18}.

Papernot et al. \cite{PATE17,ScalablePATE18} train an ensemble of `teacher' models on disjoint private data, and use the teachers as labeling oracles to train a `student' model via aggregated noisy voting by teachers. PATE ensures a strong privacy guarantee for training data, yet assumes the test dataset as public/non-private. Unlike PATE, in \sysname, model owners' training data and client input samples are private.

Abadi et al. \cite{DP-SGD16} leverage differential privacy to train deep neutral networks with built-in privacy through differentially-private stochastic gradient descent (DP-SGD). While DP-SGD offers strong privacy guarantees for training data, it is not intended for the collaborative inference setup that we explore in \sysname with private models and private inputs.

Jayaraman et al. \cite{DistribLearning18} combine differential privacy and secure multiparty computation to enable privacy-preserving collaborative training in a distributed setting. They explore output perturbation and gradient perturbation. In the output perturbation setting, parties combine local models within a secure computation and then add the differential privacy noise before revealing the model. In the collaborative inference setting, \sysname is similar to this approach because securely computed inference results are aggregated and differential privacy noise is added before the final inference result is revealed to the client. In the gradient perturbation method, the data owners collaboratively train a global model.

Lindell and Pinkas \cite{Private_data_mining} propose a system for two parties who own their private database with similar feature type, and are willing to collaborate in order to run the ID3 learning algorithm with the union of their respective databases without revealing too much about their data to each other. In the sense of multiple model owners aiming to collaboratively predict on a private input, this approach is broadly similar to \sysname.

Graepel et al. \cite{MLConfidential} study FHE-based outsourcing of execution of a machine learning algorithm while retaining confidentiality of the training and test data. Bost et al. \cite{CryptoML} design and evaluate CryptoML for privacy-preserving classification over encrypted data where both the model provider's training data and client's test data are unknown to each other. Riazi et al. \cite{Chameleon} propose Chameleon to enable two parties to jointly compute a function securely without disclosing their private input. Chameleon is based on additive secret sharing ---for linear operations and Garbled circuit ---for non-linear operations. Wagh et al. \cite{SecureNN} design and evaluate a secure computation setting for 3-party and 4-party over common DNN architectures such that no single party can learn others' private data. Giacomelli et al. \cite{CollabPredit-RF18} propose a privacy-preserving collaborative approach for Random Forest where multiple parties share their private model securely and produce encrypted prediction for a client's private input data.

\section{Conclusion}\label{sec: concl}
We presented \sysname, a system that leverages orthogonal privacy guarantees of secure multi-party computation and differential privacy to enable multiple parties that own private models to collaborate on a common ML task such as classification of a privacy-sensitive medical image owned by a client. At the core of \sysname is additive secret sharing that ensures after an iteration of a collaborative inference on a private input, model owners learn nothing about the input and clients learn nothing about the model parameters behind the collaborative inference. In addition, through differentially-private inference aggregation, \sysname limits the advance of membership inference attack by a semi-honest client. On a commodity hardware, we demonstrate the practical viability of \sysname on four datasets, for tens of model owners, with acceptable accuracy/privacy trade-off, and performance overhead amenable to speed-up using GPUs, parallelization, and more efficient secure multi-party computation protocols.

\FloatBarrier
% \vspace{1em}
\bibliographystyle{ACM-Reference-Format}
\bibliography{main}


\begin{thebibliography}{43}

%%% the Web address.

\ifx \showCODEN    \undefined \def \showCODEN     #1{\unskip}     \fi
\ifx \showDOI      \undefined \def \showDOI       #1{#1}\fi
\ifx \showISBNx    \undefined \def \showISBNx     #1{\unskip}     \fi
\ifx \showISBNxiii \undefined \def \showISBNxiii  #1{\unskip}     \fi
\ifx \showISSN     \undefined \def \showISSN      #1{\unskip}     \fi
\ifx \showLCCN     \undefined \def \showLCCN      #1{\unskip}     \fi
\ifx \shownote     \undefined \def \shownote      #1{#1}          \fi
\ifx \showarticletitle \undefined \def \showarticletitle #1{#1}   \fi
\ifx \showURL      \undefined \def \showURL       {\relax}        \fi
% invisible to TeX
\providecommand\bibfield[2]{#2}
\providecommand\bibinfo[2]{#2}
\providecommand\natexlab[1]{#1}
\providecommand\showeprint[2][]{arXiv:#2}

\bibitem[\protect\citeauthoryear{Abadi, Chu, Goodfellow, McMahan, Mironov,
  Talwar, and Zhang}{Abadi et~al\mbox{.}}{2016}]%
        {DP-SGD16}
\bibfield{author}{\bibinfo{person}{Mart{\'{\i}}n Abadi}, \bibinfo{person}{Andy
  Chu}, \bibinfo{person}{Ian~J. Goodfellow}, \bibinfo{person}{H.~Brendan
  McMahan}, \bibinfo{person}{Ilya Mironov}, \bibinfo{person}{Kunal Talwar},
  {and} \bibinfo{person}{Li Zhang}.} \bibinfo{year}{2016}\natexlab{}.
\newblock \showarticletitle{Deep Learning with Differential Privacy}. In
  \bibinfo{booktitle}{\emph{Proceedings of the 2016 {ACM} {SIGSAC} Conference
  on Computer and Communications Security, Vienna, Austria, October 24-28,
  2016}}. \bibinfo{publisher}{{ACM}}, \bibinfo{pages}{308--318}.
\newblock


\bibitem[\protect\citeauthoryear{Angelini, di~Tollo, and Roli}{Angelini
  et~al\mbox{.}}{2008}]%
        {NN-Credit-Assess}
\bibfield{author}{\bibinfo{person}{Eliana Angelini}, \bibinfo{person}{Giacomo
  di Tollo}, {and} \bibinfo{person}{Andrea Roli}.}
  \bibinfo{year}{2008}\natexlab{}.
\newblock \showarticletitle{A neural network approach for credit risk
  evaluation}.
\newblock \bibinfo{journal}{\emph{The Quarterly Review of Economics and
  Finance}} \bibinfo{volume}{48}, \bibinfo{number}{4} (\bibinfo{date}{Nov.}
  \bibinfo{year}{2008}), \bibinfo{pages}{733--755}.
\newblock
\urldef\tempurl%
\url{https://doi.org/10.1016/j.qref.2007.04.001}
\showDOI{\tempurl}


\bibitem[\protect\citeauthoryear{Attema, Mancini, Spini, Abspoel, de~Gier,
  Fehr, Veugen, van Heesch, Worm, Luca, Cramer, and Sloot}{Attema
  et~al\mbox{.}}{2018}]%
        {Decision_Making}
\bibfield{author}{\bibinfo{person}{Thomas Attema}, \bibinfo{person}{Emiliano
  Mancini}, \bibinfo{person}{Gabriele Spini}, \bibinfo{person}{Mark Abspoel},
  \bibinfo{person}{Jan de Gier}, \bibinfo{person}{Serge Fehr},
  \bibinfo{person}{Thijs Veugen}, \bibinfo{person}{Maran van Heesch},
  \bibinfo{person}{Dani{\"{e}}l Worm}, \bibinfo{person}{Andrea~De Luca},
  \bibinfo{person}{Ronald Cramer}, {and} \bibinfo{person}{Peter M.~A. Sloot}.}
  \bibinfo{year}{2018}\natexlab{}.
\newblock \showarticletitle{A New Approach to Privacy-Preserving Clinical
  Decision Support Systems for {HIV} Treatment}.
\newblock \bibinfo{journal}{\emph{CoRR}}  \bibinfo{volume}{abs/1810.01107}
  (\bibinfo{year}{2018}).
\newblock


\bibitem[\protect\citeauthoryear{Bayatbabolghani and Blanton}{Bayatbabolghani
  and Blanton}{2020}]%
        {SMPC}
\bibfield{author}{\bibinfo{person}{Fattaneh Bayatbabolghani} {and}
  \bibinfo{person}{Marina Blanton}.} \bibinfo{year}{2020}\natexlab{}.
\newblock \showarticletitle{Secure Multi-Party Computation}. In
  \bibinfo{booktitle}{\emph{PAI Workshop held in conjunction with AAAI'2020}}.
\newblock


\bibitem[\protect\citeauthoryear{Beaver}{Beaver}{2012}]%
        {SPDZ_mult}
\bibfield{author}{\bibinfo{person}{D. Beaver}.}
  \bibinfo{year}{2012}\natexlab{}.
\newblock \showarticletitle{Efficient Multiparty Protocols using Circuit
  Randomisation}.
\newblock \bibinfo{journal}{\emph{Springer}}  \bibinfo{volume}{576}
  (\bibinfo{year}{2012}).
\newblock


\bibitem[\protect\citeauthoryear{Ben{-}Or, Goldwasser, and Wigderson}{Ben{-}Or
  et~al\mbox{.}}{1988}]%
        {Secret_share1}
\bibfield{author}{\bibinfo{person}{Michael Ben{-}Or}, \bibinfo{person}{Shafi
  Goldwasser}, {and} \bibinfo{person}{Avi Wigderson}.}
  \bibinfo{year}{1988}\natexlab{}.
\newblock \showarticletitle{Completeness Theorems for Non-Cryptographic
  Fault-Tolerant Distributed Computation}. In
  \bibinfo{booktitle}{\emph{Proceedings of the 20th Annual {ACM} Symposium on
  Theory of Computing, May 2-4, 1988, Chicago, Illinois, {USA}}},
  \bibfield{editor}{\bibinfo{person}{Janos Simon}} (Ed.).
  \bibinfo{publisher}{{ACM}}, \bibinfo{pages}{1--10}.
\newblock


\bibitem[\protect\citeauthoryear{Bost, Popa, Tu, and Goldwasser}{Bost
  et~al\mbox{.}}{2015}]%
        {CryptoML}
\bibfield{author}{\bibinfo{person}{Raphael Bost}, \bibinfo{person}{Raluca~Ada
  Popa}, \bibinfo{person}{Stephen Tu}, {and} \bibinfo{person}{Shafi
  Goldwasser}.} \bibinfo{year}{2015}\natexlab{}.
\newblock \showarticletitle{Machine Learning Classification over Encrypted
  Data}. In \bibinfo{booktitle}{\emph{22nd Annual Network and Distributed
  System Security Symposium, {NDSS} 2015, San Diego, California, USA, February
  8-11, 2015}}. \bibinfo{publisher}{The Internet Society}.
\newblock


\bibitem[\protect\citeauthoryear{Chaudhuri, Monteleoni, and Sarwate}{Chaudhuri
  et~al\mbox{.}}{2011}]%
        {DP-Empirical-RM11}
\bibfield{author}{\bibinfo{person}{Kamalika Chaudhuri}, \bibinfo{person}{Claire
  Monteleoni}, {and} \bibinfo{person}{Anand~D. Sarwate}.}
  \bibinfo{year}{2011}\natexlab{}.
\newblock \showarticletitle{Differentially Private Empirical Risk
  Minimization}.
\newblock \bibinfo{journal}{\emph{J. Mach. Learn. Res.}}  \bibinfo{volume}{12}
  (\bibinfo{year}{2011}), \bibinfo{pages}{1069--1109}.
\newblock


\bibitem[\protect\citeauthoryear{Cristofaro}{Cristofaro}{2020}]%
        {survey_ML1}
\bibfield{author}{\bibinfo{person}{Emiliano~De Cristofaro}.}
  \bibinfo{year}{2020}\natexlab{}.
\newblock \showarticletitle{An Overview of Privacy in Machine Learning}.
\newblock \bibinfo{journal}{\emph{CoRR}}  \bibinfo{volume}{abs/2005.08679}
  (\bibinfo{year}{2020}).
\newblock
\showeprint[arxiv]{2005.08679}
\urldef\tempurl%
\url{https://arxiv.org/abs/2005.08679}
\showURL{%
\tempurl}


\bibitem[\protect\citeauthoryear{Dahl, Yu, Deng, and Acero}{Dahl
  et~al\mbox{.}}{2012}]%
        {DL-Speech2012}
\bibfield{author}{\bibinfo{person}{G.~E. Dahl}, \bibinfo{person}{Dong Yu},
  \bibinfo{person}{Li Deng}, {and} \bibinfo{person}{A. Acero}.}
  \bibinfo{year}{2012}\natexlab{}.
\newblock \showarticletitle{Context-Dependent Pre-Trained Deep Neural Networks
  for Large-Vocabulary Speech Recognition}.
\newblock \bibinfo{journal}{\emph{{IEEE} Transactions on Audio, Speech, and
  Language Processing}} \bibinfo{volume}{20}, \bibinfo{number}{1}
  (\bibinfo{date}{Jan.} \bibinfo{year}{2012}), \bibinfo{pages}{30--42}.
\newblock
\urldef\tempurl%
\url{https://doi.org/10.1109/tasl.2011.2134090}
\showDOI{\tempurl}


\bibitem[\protect\citeauthoryear{Damg{\aa}rd, Pastro, Smart, and
  Zakarias}{Damg{\aa}rd et~al\mbox{.}}{2012}]%
        {SPDZ_protocol}
\bibfield{author}{\bibinfo{person}{Ivan Damg{\aa}rd}, \bibinfo{person}{Valerio
  Pastro}, \bibinfo{person}{Nigel~P. Smart}, {and} \bibinfo{person}{Sarah
  Zakarias}.} \bibinfo{year}{2012}\natexlab{}.
\newblock \showarticletitle{Multiparty Computation from Somewhat Homomorphic
  Encryption}. In \bibinfo{booktitle}{\emph{Advances in Cryptology - {CRYPTO}
  2012 - 32nd Annual Cryptology Conference, Santa Barbara, CA, USA, August
  19-23, 2012. Proceedings}} \emph{(\bibinfo{series}{Lecture Notes in Computer
  Science})}, \bibfield{editor}{\bibinfo{person}{Reihaneh Safavi{-}Naini} {and}
  \bibinfo{person}{Ran Canetti}} (Eds.), Vol.~\bibinfo{volume}{7417}.
  \bibinfo{publisher}{Springer}, \bibinfo{pages}{643--662}.
\newblock


\bibitem[\protect\citeauthoryear{Dwork, McSherry, Nissim, and Smith}{Dwork
  et~al\mbox{.}}{2006}]%
        {Differential_Privacy}
\bibfield{author}{\bibinfo{person}{Cynthia Dwork}, \bibinfo{person}{Frank
  McSherry}, \bibinfo{person}{Kobbi Nissim}, {and} \bibinfo{person}{Adam~D.
  Smith}.} \bibinfo{year}{2006}\natexlab{}.
\newblock \showarticletitle{Calibrating Noise to Sensitivity in Private Data
  Analysis}. In \bibinfo{booktitle}{\emph{Theory of Cryptography, Third Theory
  of Cryptography Conference, {TCC} 2006, New York, NY, USA, March 4-7, 2006,
  Proceedings}} \emph{(\bibinfo{series}{Lecture Notes in Computer Science})},
  \bibfield{editor}{\bibinfo{person}{Shai Halevi} {and} \bibinfo{person}{Tal
  Rabin}} (Eds.), Vol.~\bibinfo{volume}{3876}. \bibinfo{publisher}{Springer},
  \bibinfo{pages}{265--284}.
\newblock


\bibitem[\protect\citeauthoryear{Fredrikson, Jha, and Ristenpart}{Fredrikson
  et~al\mbox{.}}{2015}]%
        {model-inversion}
\bibfield{author}{\bibinfo{person}{Matt Fredrikson}, \bibinfo{person}{Somesh
  Jha}, {and} \bibinfo{person}{Thomas Ristenpart}.}
  \bibinfo{year}{2015}\natexlab{}.
\newblock \showarticletitle{Model Inversion Attacks that Exploit Confidence
  Information and Basic Countermeasures}. In
  \bibinfo{booktitle}{\emph{Proceedings of the 22nd {ACM} {SIGSAC} Conference
  on Computer and Communications Security, Denver, CO, USA, October 12-16,
  2015}}. \bibinfo{pages}{1322--1333}.
\newblock


\bibitem[\protect\citeauthoryear{Gao, Wang, Tan, Zhu, Zhang, Fessler,
  Vermeulen, and Wang}{Gao et~al\mbox{.}}{2019}]%
        {DeepCC2019}
\bibfield{author}{\bibinfo{person}{Feng Gao}, \bibinfo{person}{Wei Wang},
  \bibinfo{person}{Miaomiao Tan}, \bibinfo{person}{Lina Zhu},
  \bibinfo{person}{Yuchen Zhang}, \bibinfo{person}{Evelyn Fessler},
  \bibinfo{person}{Louis Vermeulen}, {and} \bibinfo{person}{Xin Wang}.}
  \bibinfo{year}{2019}\natexlab{}.
\newblock \showarticletitle{{DeepCC}: a novel deep learning-based framework for
  cancer molecular subtype classification}.
\newblock \bibinfo{journal}{\emph{Oncogenesis}} \bibinfo{volume}{8},
  \bibinfo{number}{9} (\bibinfo{date}{Aug.} \bibinfo{year}{2019}).
\newblock
\urldef\tempurl%
\url{https://doi.org/10.1038/s41389-019-0157-8}
\showDOI{\tempurl}


\bibitem[\protect\citeauthoryear{Giacomelli, Jha, Kleiman, Page, and
  Yoon}{Giacomelli et~al\mbox{.}}{2018}]%
        {CollabPredit-RF18}
\bibfield{author}{\bibinfo{person}{Irene Giacomelli}, \bibinfo{person}{Somesh
  Jha}, \bibinfo{person}{Ross Kleiman}, \bibinfo{person}{David Page}, {and}
  \bibinfo{person}{Kyonghwan Yoon}.} \bibinfo{year}{2018}\natexlab{}.
\newblock \showarticletitle{Privacy-Preserving Collaborative Prediction using
  Random Forests}.
\newblock \bibinfo{journal}{\emph{CoRR}}  \bibinfo{volume}{abs/1811.08695}
  (\bibinfo{year}{2018}).
\newblock
\showeprint[arxiv]{1811.08695}
\urldef\tempurl%
\url{http://arxiv.org/abs/1811.08695}
\showURL{%
\tempurl}


\bibitem[\protect\citeauthoryear{Gilad{-}Bachrach, Dowlin, Laine, Lauter,
  Naehrig, and Wernsing}{Gilad{-}Bachrach et~al\mbox{.}}{2016}]%
        {CryptoNets16}
\bibfield{author}{\bibinfo{person}{Ran Gilad{-}Bachrach},
  \bibinfo{person}{Nathan Dowlin}, \bibinfo{person}{Kim Laine},
  \bibinfo{person}{Kristin~E. Lauter}, \bibinfo{person}{Michael Naehrig}, {and}
  \bibinfo{person}{John Wernsing}.} \bibinfo{year}{2016}\natexlab{}.
\newblock \showarticletitle{CryptoNets: Applying Neural Networks to Encrypted
  Data with High Throughput and Accuracy}. In
  \bibinfo{booktitle}{\emph{Proceedings of the 33nd International Conference on
  Machine Learning, {ICML} 2016, New York City, NY, USA, June 19-24, 2016}}
  \emph{(\bibinfo{series}{{JMLR} Workshop and Conference Proceedings})},
  \bibfield{editor}{\bibinfo{person}{Maria{-}Florina Balcan} {and}
  \bibinfo{person}{Kilian~Q. Weinberger}} (Eds.), Vol.~\bibinfo{volume}{48}.
  \bibinfo{publisher}{JMLR.org}, \bibinfo{pages}{201--210}.
\newblock


\bibitem[\protect\citeauthoryear{Graepel, Lauter, and Naehrig}{Graepel
  et~al\mbox{.}}{2012}]%
        {MLConfidential}
\bibfield{author}{\bibinfo{person}{Thore Graepel}, \bibinfo{person}{Kristin~E.
  Lauter}, {and} \bibinfo{person}{Michael Naehrig}.}
  \bibinfo{year}{2012}\natexlab{}.
\newblock \showarticletitle{{ML} Confidential: Machine Learning on Encrypted
  Data}. In \bibinfo{booktitle}{\emph{Information Security and Cryptology -
  {ICISC} 2012 - 15th International Conference, Seoul, Korea, November 28-30,
  2012, Revised Selected Papers}} \emph{(\bibinfo{series}{Lecture Notes in
  Computer Science})}, \bibfield{editor}{\bibinfo{person}{Taekyoung Kwon},
  \bibinfo{person}{Mun{-}Kyu Lee}, {and} \bibinfo{person}{Daesung Kwon}}
  (Eds.), Vol.~\bibinfo{volume}{7839}. \bibinfo{publisher}{Springer},
  \bibinfo{pages}{1--21}.
\newblock


\bibitem[\protect\citeauthoryear{Haagh, Karbyshev, Oechsner, Spitters, and
  Strub}{Haagh et~al\mbox{.}}{2018}]%
        {Proof_SMPC}
\bibfield{author}{\bibinfo{person}{Helene Haagh}, \bibinfo{person}{Aleksandr
  Karbyshev}, \bibinfo{person}{Sabine Oechsner}, \bibinfo{person}{Bas
  Spitters}, {and} \bibinfo{person}{Pierre{-}Yves Strub}.}
  \bibinfo{year}{2018}\natexlab{}.
\newblock \showarticletitle{Computer-Aided Proofs for Multiparty Computation
  with Active Security}. In \bibinfo{booktitle}{\emph{31st {IEEE} Computer
  Security Foundations Symposium, {CSF} 2018, Oxford, United Kingdom, July
  9-12, 2018}}. \bibinfo{publisher}{{IEEE} Computer Society},
  \bibinfo{pages}{119--131}.
\newblock


\bibitem[\protect\citeauthoryear{Jacobson}{Jacobson}{1957}]%
        {Q_ring}
\bibfield{author}{\bibinfo{person}{Nathan Jacobson}.}
  \bibinfo{year}{1957}\natexlab{}.
\newblock \showarticletitle{Structure of Rings}. In
  \bibinfo{booktitle}{\emph{Bull. Amer. Math. Soc. 63 (1957)}}.
  \bibinfo{pages}{46--50.}
\newblock


\bibitem[\protect\citeauthoryear{Jayaraman, Wang, Evans, and Gu}{Jayaraman
  et~al\mbox{.}}{2018}]%
        {DistribLearning18}
\bibfield{author}{\bibinfo{person}{Bargav Jayaraman}, \bibinfo{person}{Lingxiao
  Wang}, \bibinfo{person}{David Evans}, {and} \bibinfo{person}{Quanquan Gu}.}
  \bibinfo{year}{2018}\natexlab{}.
\newblock \showarticletitle{Distributed Learning without Distress:
  Privacy-Preserving Empirical Risk Minimization}. In
  \bibinfo{booktitle}{\emph{Advances in Neural Information Processing Systems
  31: Annual Conference on Neural Information Processing Systems 2018, NeurIPS
  2018, 3-8 December 2018, Montr{\'{e}}al, Canada}}.
  \bibinfo{pages}{6346--6357}.
\newblock


\bibitem[\protect\citeauthoryear{Joan~Feigenbaum and
  Saint-Jean}{Joan~Feigenbaum and Saint-Jean}{2004}]%
        {Secure_Auction}
\bibfield{author}{\bibinfo{person}{Raphael~Ryger Joan~Feigenbaum, Benny~Pinkas}
  {and} \bibinfo{person}{Felipe Saint-Jean}.} \bibinfo{year}{2004}\natexlab{}.
\newblock \showarticletitle{Secure Computation of Surveys}. In
  \bibinfo{booktitle}{\emph{EU Workshop on Secure Multiparty Protocols, 2004}}.
\newblock


\bibitem[\protect\citeauthoryear{Johnson, Pollard, Shen, Lehman, Feng,
  Ghassemi, Moody, Szolovits, Anthony~Celi, and Mark}{Johnson
  et~al\mbox{.}}{2016}]%
        {MIMIC}
\bibfield{author}{\bibinfo{person}{Alistair~E.W. Johnson},
  \bibinfo{person}{Tom~J. Pollard}, \bibinfo{person}{Lu Shen},
  \bibinfo{person}{Li-wei~H. Lehman}, \bibinfo{person}{Mengling Feng},
  \bibinfo{person}{Mohammad Ghassemi}, \bibinfo{person}{Benjamin Moody},
  \bibinfo{person}{Peter Szolovits}, \bibinfo{person}{Leo Anthony~Celi}, {and}
  \bibinfo{person}{Roger~G. Mark}.} \bibinfo{year}{2016}\natexlab{}.
\newblock \showarticletitle{MIMIC-III, a freely accessible critical care
  database}.
\newblock \bibinfo{journal}{\emph{Scientific Data}} \bibinfo{volume}{3},
  \bibinfo{number}{1} (\bibinfo{date}{24 May} \bibinfo{year}{2016}),
  \bibinfo{pages}{160035}.
\newblock
\showISSN{2052-4463}
\urldef\tempurl%
\url{https://doi.org/10.1038/sdata.2016.35}
\showDOI{\tempurl}


\bibitem[\protect\citeauthoryear{Kaggle}{Kaggle}{2020}]%
        {IDC}
\bibfield{author}{\bibinfo{person}{Kaggle}.} \bibinfo{year}{2020}\natexlab{}.
\newblock \bibinfo{title}{Invasive Ductal Carcinoma Dataset}.
\newblock \bibinfo{howpublished}{\url{https://www.kaggle.com/paultimothymooney/
  breast-histopathology-images}}.
\newblock


\bibitem[\protect\citeauthoryear{Krizhevsky, Sutskever, and Hinton}{Krizhevsky
  et~al\mbox{.}}{2017}]%
        {ImageNet}
\bibfield{author}{\bibinfo{person}{Alex Krizhevsky}, \bibinfo{person}{Ilya
  Sutskever}, {and} \bibinfo{person}{Geoffrey~E. Hinton}.}
  \bibinfo{year}{2017}\natexlab{}.
\newblock \showarticletitle{ImageNet classification with deep convolutional
  neural networks}.
\newblock \bibinfo{journal}{\emph{Commun. {ACM}}} \bibinfo{volume}{60},
  \bibinfo{number}{6} (\bibinfo{year}{2017}), \bibinfo{pages}{84--90}.
\newblock


\bibitem[\protect\citeauthoryear{LeCun, Cortes, and Burges}{LeCun
  et~al\mbox{.}}{2020}]%
        {MNIST}
\bibfield{author}{\bibinfo{person}{Yan LeCun}, \bibinfo{person}{Corinna
  Cortes}, {and} \bibinfo{person}{Christopher~J.C. Burges}.}
  \bibinfo{year}{2020}\natexlab{}.
\newblock \bibinfo{title}{The MNIST Database of Handwritten Digits}.
\newblock \bibinfo{howpublished}{\url{http://yann.lecun.com/exdb/mnist/}}.
\newblock


\bibitem[\protect\citeauthoryear{Lindell and Pinkas}{Lindell and
  Pinkas}{2000}]%
        {Private_data_mining}
\bibfield{author}{\bibinfo{person}{Yehuda Lindell} {and} \bibinfo{person}{Benny
  Pinkas}.} \bibinfo{year}{2000}\natexlab{}.
\newblock \showarticletitle{Privacy Preserving Data Mining}. In
  \bibinfo{booktitle}{\emph{Advances in Cryptology - {CRYPTO} 2000, 20th Annual
  International Cryptology Conference, Santa Barbara, California, USA, August
  20-24, 2000, Proceedings}} \emph{(\bibinfo{series}{Lecture Notes in Computer
  Science})}, \bibfield{editor}{\bibinfo{person}{Mihir Bellare}} (Ed.),
  Vol.~\bibinfo{volume}{1880}. \bibinfo{publisher}{Springer},
  \bibinfo{pages}{36--54}.
\newblock


\bibitem[\protect\citeauthoryear{Liu, Juuti, Lu, and Asokan}{Liu
  et~al\mbox{.}}{2017}]%
        {MiniONN17}
\bibfield{author}{\bibinfo{person}{Jian Liu}, \bibinfo{person}{Mika Juuti},
  \bibinfo{person}{Yao Lu}, {and} \bibinfo{person}{N. Asokan}.}
  \bibinfo{year}{2017}\natexlab{}.
\newblock \showarticletitle{Oblivious Neural Network Predictions via MiniONN
  Transformations}. In \bibinfo{booktitle}{\emph{Proceedings of the 2017 {ACM}
  {SIGSAC} Conference on Computer and Communications Security, {CCS} 2017,
  Dallas, TX, USA, October 30 - November 03, 2017}},
  \bibfield{editor}{\bibinfo{person}{Bhavani~M. Thuraisingham},
  \bibinfo{person}{David Evans}, \bibinfo{person}{Tal Malkin}, {and}
  \bibinfo{person}{Dongyan Xu}} (Eds.). \bibinfo{publisher}{{ACM}},
  \bibinfo{pages}{619--631}.
\newblock


\bibitem[\protect\citeauthoryear{Mohassel and Zhang}{Mohassel and
  Zhang}{2017}]%
        {SecureML17}
\bibfield{author}{\bibinfo{person}{Payman Mohassel} {and}
  \bibinfo{person}{Yupeng Zhang}.} \bibinfo{year}{2017}\natexlab{}.
\newblock \showarticletitle{SecureML: {A} System for Scalable
  Privacy-Preserving Machine Learning}. In \bibinfo{booktitle}{\emph{2017
  {IEEE} Symposium on Security and Privacy, {SP} 2017, San Jose, CA, USA, May
  22-26, 2017}}. \bibinfo{publisher}{{IEEE} Computer Society},
  \bibinfo{pages}{19--38}.
\newblock


\bibitem[\protect\citeauthoryear{Nair, Binu, and Kumar}{Nair
  et~al\mbox{.}}{2015}]%
        {Secure_Voting}
\bibfield{author}{\bibinfo{person}{Divya~G. Nair}, \bibinfo{person}{V.~P.
  Binu}, {and} \bibinfo{person}{G.~Santhosh Kumar}.}
  \bibinfo{year}{2015}\natexlab{}.
\newblock \showarticletitle{An Improved E-voting scheme using Secret Sharing
  based Secure Multi-party Computation}.
\newblock \bibinfo{journal}{\emph{CoRR}}  \bibinfo{volume}{abs/1502.07469}
  (\bibinfo{year}{2015}).
\newblock


\bibitem[\protect\citeauthoryear{OpenMined}{OpenMined}{2020}]%
        {PySyft}
\bibfield{author}{\bibinfo{person}{OpenMined}.}
  \bibinfo{year}{2020}\natexlab{}.
\newblock \bibinfo{title}{PySyft}.
\newblock
  \bibinfo{howpublished}{\url{https://github.com/OpenMined/PySyft/blob/master/syft/frameworks/torch/mpc/securenn.py}}.
\newblock


\bibitem[\protect\citeauthoryear{Papernot, Abadi, Erlingsson, Goodfellow, and
  Talwar}{Papernot et~al\mbox{.}}{2017}]%
        {PATE17}
\bibfield{author}{\bibinfo{person}{Nicolas Papernot},
  \bibinfo{person}{Mart{\'{\i}}n Abadi}, \bibinfo{person}{{\'{U}}lfar
  Erlingsson}, \bibinfo{person}{Ian~J. Goodfellow}, {and}
  \bibinfo{person}{Kunal Talwar}.} \bibinfo{year}{2017}\natexlab{}.
\newblock \showarticletitle{Semi-supervised Knowledge Transfer for Deep
  Learning from Private Training Data}. In \bibinfo{booktitle}{\emph{5th
  International Conference on Learning Representations, {ICLR} 2017}}.
\newblock


\bibitem[\protect\citeauthoryear{Papernot, McDaniel, Sinha, and
  Wellman}{Papernot et~al\mbox{.}}{2018a}]%
        {SP-SoK18}
\bibfield{author}{\bibinfo{person}{Nicolas Papernot},
  \bibinfo{person}{Patrick~D. McDaniel}, \bibinfo{person}{Arunesh Sinha}, {and}
  \bibinfo{person}{Michael~P. Wellman}.} \bibinfo{year}{2018}\natexlab{a}.
\newblock \showarticletitle{SoK: Security and Privacy in Machine Learning}. In
  \bibinfo{booktitle}{\emph{2018 {IEEE} European Symposium on Security and
  Privacy, EuroS{\&}P 2018, London, United Kingdom, April 24-26, 2018}}.
  \bibinfo{pages}{399--414}.
\newblock


\bibitem[\protect\citeauthoryear{Papernot, Song, Mironov, Raghunathan, Talwar,
  and Erlingsson}{Papernot et~al\mbox{.}}{2018b}]%
        {ScalablePATE18}
\bibfield{author}{\bibinfo{person}{Nicolas Papernot}, \bibinfo{person}{Shuang
  Song}, \bibinfo{person}{Ilya Mironov}, \bibinfo{person}{Ananth Raghunathan},
  \bibinfo{person}{Kunal Talwar}, {and} \bibinfo{person}{{\'{U}}lfar
  Erlingsson}.} \bibinfo{year}{2018}\natexlab{b}.
\newblock \showarticletitle{Scalable Private Learning with {PATE}}. In
  \bibinfo{booktitle}{\emph{6th International Conference on Learning
  Representations, {ICLR} 2018, Vancouver, BC, Canada, April 30 - May 3, 2018,
  Conference Track Proceedings}}.
\newblock


\bibitem[\protect\citeauthoryear{Rahman, Rahman, Lagani{\`{e}}re, and
  Mohammed}{Rahman et~al\mbox{.}}{2018}]%
        {MIA_DP}
\bibfield{author}{\bibinfo{person}{Md.~Atiqur Rahman}, \bibinfo{person}{Tanzila
  Rahman}, \bibinfo{person}{Robert Lagani{\`{e}}re}, {and}
  \bibinfo{person}{Noman Mohammed}.} \bibinfo{year}{2018}\natexlab{}.
\newblock \showarticletitle{Membership Inference Attack against Differentially
  Private Deep Learning Model}.
\newblock \bibinfo{journal}{\emph{Trans. Data Priv.}} \bibinfo{volume}{11},
  \bibinfo{number}{1} (\bibinfo{year}{2018}), \bibinfo{pages}{61--79}.
\newblock


\bibitem[\protect\citeauthoryear{Riazi, Weinert, Tkachenko, Songhori,
  Schneider, and Koushanfar}{Riazi et~al\mbox{.}}{2018}]%
        {Chameleon}
\bibfield{author}{\bibinfo{person}{M.~Sadegh Riazi}, \bibinfo{person}{Christian
  Weinert}, \bibinfo{person}{Oleksandr Tkachenko}, \bibinfo{person}{Ebrahim~M.
  Songhori}, \bibinfo{person}{Thomas Schneider}, {and} \bibinfo{person}{Farinaz
  Koushanfar}.} \bibinfo{year}{2018}\natexlab{}.
\newblock \showarticletitle{Chameleon: {A} Hybrid Secure Computation Framework
  for Machine Learning Applications}. In \bibinfo{booktitle}{\emph{Proceedings
  of the 2018 on Asia Conference on Computer and Communications Security,
  AsiaCCS 2018, Incheon, Republic of Korea, June 04-08, 2018}},
  \bibfield{editor}{\bibinfo{person}{Jong Kim}, \bibinfo{person}{Gail{-}Joon
  Ahn}, \bibinfo{person}{Seungjoo Kim}, \bibinfo{person}{Yongdae Kim},
  \bibinfo{person}{Javier L{\'{o}}pez}, {and} \bibinfo{person}{Taesoo Kim}}
  (Eds.). \bibinfo{publisher}{{ACM}}, \bibinfo{pages}{707--721}.
\newblock


\bibitem[\protect\citeauthoryear{Sallab, Abdou, Perot, and Yogamani}{Sallab
  et~al\mbox{.}}{2017}]%
        {DL-autnonmous17}
\bibfield{author}{\bibinfo{person}{Ahmad~El Sallab}, \bibinfo{person}{Mohammed
  Abdou}, \bibinfo{person}{Etienne Perot}, {and} \bibinfo{person}{Senthil~Kumar
  Yogamani}.} \bibinfo{year}{2017}\natexlab{}.
\newblock \showarticletitle{Deep Reinforcement Learning framework for
  Autonomous Driving}.
\newblock \bibinfo{journal}{\emph{CoRR}}  \bibinfo{volume}{abs/1704.02532}
  (\bibinfo{year}{2017}).
\newblock


\bibitem[\protect\citeauthoryear{Shamir}{Shamir}{1979}]%
        {Secret_share2}
\bibfield{author}{\bibinfo{person}{Adi Shamir}.}
  \bibinfo{year}{1979}\natexlab{}.
\newblock \showarticletitle{How to Share a Secret}.
\newblock \bibinfo{journal}{\emph{Commun. {ACM}}} \bibinfo{volume}{22},
  \bibinfo{number}{11} (\bibinfo{year}{1979}), \bibinfo{pages}{612--613}.
\newblock


\bibitem[\protect\citeauthoryear{Shokri, Stronati, Song, and Shmatikov}{Shokri
  et~al\mbox{.}}{2017}]%
        {MIA}
\bibfield{author}{\bibinfo{person}{Reza Shokri}, \bibinfo{person}{Marco
  Stronati}, \bibinfo{person}{Congzheng Song}, {and} \bibinfo{person}{Vitaly
  Shmatikov}.} \bibinfo{year}{2017}\natexlab{}.
\newblock \showarticletitle{Membership Inference Attacks Against Machine
  Learning Models}. In \bibinfo{booktitle}{\emph{2017 {IEEE} Symposium on
  Security and Privacy, {SP} 2017, San Jose, CA, USA, May 22-26, 2017}}.
  \bibinfo{publisher}{{IEEE} Computer Society}, \bibinfo{pages}{3--18}.
\newblock


\bibitem[\protect\citeauthoryear{Tram{\`{e}}r, Zhang, Juels, Reiter, and
  Ristenpart}{Tram{\`{e}}r et~al\mbox{.}}{2016}]%
        {model-stealing16}
\bibfield{author}{\bibinfo{person}{Florian Tram{\`{e}}r}, \bibinfo{person}{Fan
  Zhang}, \bibinfo{person}{Ari Juels}, \bibinfo{person}{Michael~K. Reiter},
  {and} \bibinfo{person}{Thomas Ristenpart}.} \bibinfo{year}{2016}\natexlab{}.
\newblock \showarticletitle{Stealing Machine Learning Models via Prediction
  APIs}. In \bibinfo{booktitle}{\emph{25th {USENIX} Security Symposium,
  {USENIX} Security 16, Austin, TX, USA, August 10-12, 2016}}.
  \bibinfo{pages}{601--618}.
\newblock


\bibitem[\protect\citeauthoryear{Wagh, Gupta, and Chandran}{Wagh
  et~al\mbox{.}}{2018}]%
        {SecureNN}
\bibfield{author}{\bibinfo{person}{Sameer Wagh}, \bibinfo{person}{Divya Gupta},
  {and} \bibinfo{person}{Nishanth Chandran}.} \bibinfo{year}{2018}\natexlab{}.
\newblock \showarticletitle{SecureNN: Efficient and Private Neural Network
  Training}.
\newblock \bibinfo{journal}{\emph{{IACR} Cryptol. ePrint Arch.}}
  \bibinfo{volume}{2018} (\bibinfo{year}{2018}), \bibinfo{pages}{442}.
\newblock


\bibitem[\protect\citeauthoryear{Wagh, Gupta, and Chandran}{Wagh
  et~al\mbox{.}}{2019}]%
        {SecureNN_1}
\bibfield{author}{\bibinfo{person}{Sameer Wagh}, \bibinfo{person}{Divya Gupta},
  {and} \bibinfo{person}{Nishanth Chandran}.} \bibinfo{year}{2019}\natexlab{}.
\newblock \showarticletitle{SecureNN: 3-Party Secure Computation for Neural
  Network Training}.
\newblock \bibinfo{journal}{\emph{Proc. Priv. Enhancing Technol.}}
  \bibinfo{volume}{2019}, \bibinfo{number}{3} (\bibinfo{year}{2019}),
  \bibinfo{pages}{26--49}.
\newblock


\bibitem[\protect\citeauthoryear{Xiao, Rasul, and Vollgraf}{Xiao
  et~al\mbox{.}}{2017}]%
        {Fashion-MNIST}
\bibfield{author}{\bibinfo{person}{Han Xiao}, \bibinfo{person}{Kashif Rasul},
  {and} \bibinfo{person}{Roland Vollgraf}.} \bibinfo{year}{2017}\natexlab{}.
\newblock \bibinfo{booktitle}{\emph{Fashion-MNIST: a Novel Image Dataset for
  Benchmarking Machine Learning Algorithms}}.
\newblock
\showeprint[arXiv]{cs.LG/cs.LG/1708.07747}


\bibitem[\protect\citeauthoryear{Yao}{Yao}{1986}]%
        {GC}
\bibfield{author}{\bibinfo{person}{Andrew~Chi{-}Chih Yao}.}
  \bibinfo{year}{1986}\natexlab{}.
\newblock \showarticletitle{How to Generate and Exchange Secrets (Extended
  Abstract)}. In \bibinfo{booktitle}{\emph{27th Annual Symposium on Foundations
  of Computer Science, Toronto, Canada, 27-29 October 1986}}.
  \bibinfo{publisher}{{IEEE} Computer Society}, \bibinfo{pages}{162--167}.
\newblock


\end{thebibliography}
\section{Appendix}\label{sec: Appendix}
\subsection{Model Architectures Used for Evaluation}\label{subsec: model-architectures}
\begin{centering}
\framebox{

\parbox[t][3.6cm]{7.7 cm}{

\addvspace{0.2cm} \centering 

\makebox{\large{Model Architecture of MNIST and Fashion-MNIST}}\\
\vspace{.1cm}
Number of input neurons $=28\times 28$\\
Number of output neurons $=10$\\
Number of hidden layers $l=2$\\
Number of neurons in $1^{st}$ layer =$128$\\
Number of neurons in $2^{nd}$ layer =$64$\\
Activation Function: \textbf{ReLu } \\
%Learning Rate=.001\\
%Number of epochs =100\\
Error estimation method: MSE

} 

}

\framebox{
\parbox[t][3.1cm]{7.7 cm}{

\addvspace{0.2cm} \centering 

\makebox{\large{Model Architecture of IDC Regular }}\\
\vspace{.1cm}
Number of input neurons $=7500$\\
Number of output neurons $=2$\\
Number of hidden layers $l=1$\\
Number of neurons in hidden layer =$500$\\
Activation Function: \textbf{ReLu }\\
%Learning Rate=.001\\
%Number of epochs =80\\
Error estimation method: MSE

} 

}

\framebox{
\parbox[t][3.1cm]{7.7 cm}{

\addvspace{0.2cm} \centering 

\makebox{\large{Model Architecture of MIMIC Critical Care}}\\
\vspace{.1cm}
Number of input neurons $=30$\\
Number of output neurons $=4$\\
Number of hidden layers $l=1$\\
Number of neurons in hidden layer =$500$\\
Activation Function: \textbf{ReLu }\\
%Learning Rate=.001\\
%Number of epochs =80\\
Error estimation method: MSE

} 

}

\end{centering}

\subsection{Illustration of Additive Secret Sharing} \label{subsec: secret-sharing-proof}
For additive secret sharing, first we have a finite field ${0, 1, 2,...... Q-1}$ where all the secret-shares belong where $Q$ is a very large prime number. Now, suppose we split up $x$ into $i$ shares such that $x$  = $x_1,x_2,x_3, ..., x_{i-1}$ and the $i^{th}$ share will be $x_i=x-(x_1 + x_2+, ..., +x_i)$,  and the reconstruction will be $x=\sum_{k=1}^{i} x_k$. \cite{Proof_SMPC}. 
For two workers A and B, input value $x_1$ is  split into two secret-shares and sent to A and B as follows: 

\vspace{-.25in}
\begin{align*}
    A: x_1^a=Random (-Q,Q)\\
    B: x_1^b=x_1-x_1^a
\end{align*}

The reconstruction is done using the following procedure: 
\begin{align*}
    x_1=(x_1^a+x_1^b) \quad Mod \quad(Q)
\end{align*}
The reconstruction can be proved with as follows: 
\vspace{-.25in}

\begin{align*}
(x_1^a+x_1^b) \%Q &= (x_1^a+((x_1-x_1^a)\%Q))\%Q \\
&=(x_1^a\%Q) + (x_1\%Q)\%Q -(x_1^a\%Q)\%Q\\
&=(x_1^a\%Q) + (x_1\%Q)\%Q -(x_1^a\%Q)\\
&= (x_1\%Q) \quad [(a\%n1)\%n1=a\%n1]\\
&= x_1  \quad[Q >> x, Q: large \quad prime]
\end{align*}

For summation, suppose we want to compute  $x+y$ via secret sharing. We divide them using Equation \ref{eq:share-a} and \ref{eq:share-b} as follows: 
\vspace{-.25in}

\begin{align*}
    A: x^a = Random (-Q,Q), y^a = Random(-Q,Q)\\
    B: x^b = (x-x^a) \quad mod \quad Q, y^b = (Y-Y^a) \quad mod \quad Q
\end{align*}

Now,  the system sends $x^a$ and $y^a$ to worker $A$ and $x^b$ and $y^b$ to worker $B$. The summation operation on each worker side will be: 
\vspace{-.25in}

\begin{align*}
    A: x^a +y^a \\
    B: x^b+y^b
\end{align*}
The actual sum is reconstructed using Equation \ref{eq:combine-shares} as follows: 
\begin{align*}
    sum=(x^a +y^a+x^b+y^b) \quad Mod \quad(Q)
\end{align*}

This reconstruction can be proved as follows: 
\vspace{-.25in}

\begin{align*}
&((x^a +y^a+x^b+y^b))\%Q \\
&=(x^a\%Q)+(y^a\%Q) -(x^a\%Q)\%Q-(x^a\%Q)\%Q+(x\%Q)\%Q+(y\%Q)\%Q\\
&=(x^a\%Q)+(y^a\%Q) -(x^a\%Q)-(y^a\%Q)+(x\%Q)+(y\%Q)\\
&= (x\%Q)+(y\%Q) \quad [(a\%n1)\%n1=a\%n1]\\
&= x+y  \quad[Q >> x,y]\\
\end{align*}
Since Q is a large prime compared to $x$ and $Y$, the above illustration proves that additive secret sharing reconstructs $x+y$.

\subsection{Illustration of SPDZ}\label{subsec: spdz-proof}
For two numbers $x$ and $y$, our goal is to compute $x*y$ securely. We divide them using Equation \ref{eq:share-a} and \ref{eq:share-b} as follows: 
\vspace{-.25in}

\begin{align*}
    A: x^a = Random (-Q,Q), y^a = Random(-Q,Q)\\
     B: x^b = (x-x^a) \quad mod \quad Q, y^b = (Y-Y^a) \quad mod \quad Q
\end{align*}

The multiplication operation will be: 
\vspace{-.25in}

\begin{align*}
(x^a+x^b)(y^a+y^b)=x^a y^a +x^a y^b + x^b y^a +x^b y^b.
\end{align*}
 
We can observe from above that to calculate $x^a y^b $ and $x^b y^a$ Worker A and Worker B need each other's share, which conflicts with the privacy assumptions of our approach. To resolve the conflict, the SPDZ protocol \cite{SPDZ_mult} is used for performing secret sharing multiplication. For this technique, independent of $x$ and $y$, values called `multiplication triplets' are generated in offline phase. In this phase, three numbers $q1$, $q2$ and $q3$ are generated and shared with two parties ahead of running the protocol. The numbers, $q1$, $q2$ and $q3$ are generated as follows:
\begin{align*}
  q1&=Random (-Q,Q)\\
  q2&=Random (-Q,Q)\\
  q3&=(q1\times q2) \quad Mod \quad (Q)
\end{align*}
 % here I have to write the background conceptual overview of SPDZ 
 Now, these $q1,q2$ and $q3$ parameters are secretly shared with the workers and the workers will compute: 
 \vspace{-.25in}

 \begin{align*}
  \alpha^a&=x^a-q^a_1\\
  \beta^a&=y^a-q^a_1
\end{align*}
And for worker B:
 \begin{align*}
  \alpha^b&=x^b-q^b_1\\
  \beta^b&=y^b-q^b_1
\end{align*}

And reconstruct them using Equation \ref{eq:combine-shares} and compute $\alpha$ and $\beta$. This two parameters will be public to worker A and Worker B. Now the intermediate results of multiplication of $x$ and $y$ will be,
\vspace{-.25in}

 \begin{align*}
  Worker A: (xy)^a=q^a_3+(\alpha^a)q^a_2+(\beta^a)q^a_1\\
  Worker B: (xy)^b=q^b_3+(\alpha^b)q^b_2+(\beta^b)q^b_1
\end{align*}
If we reconstruct $(xy)^a$ and $(xy)^b$ using Equation \ref{eq:combine-shares} the real multiplication value of $xy$ will be obtained. 
%\cite{SK1}

\end{document}